\documentstyle{mn}
\input epsfig.sty
\input psfig.sty
\def\sles{\lower2pt\hbox{$\buildrel {\scriptstyle <}
   \over {\scriptstyle\sim}$}}
\def\sgreat{\lower2pt\hbox{$\buildrel {\scriptstyle >}
   \over {\scriptstyle\sim}$}}
\title{A class of exact MHD models for astrophysical jets}

\author[N. ~Vlahakis \& K. ~Tsinganos]
 {N.~Vlahakis$^1$\thanks{Email:  vlahakis@physics.uch.gr; 
tsingan@physics.uch.gr} and K.~Tsinganos$^{1,2,\star}$\\
 $^1$Department of Physics, University of Crete, GR-710 03 Heraklion, Crete, 
 GREECE\\
 $^2$Foundation for Research and Technology Hellas (FORTH), 
 GR-711 10 Heraklion, Crete, GREECE\\
 }
\date{Accepted February 8, 1999. Received December 8, 1998; in original form 1998 July 21} 
\pagerange{\pageref{firstpage}--\pageref{lastpage}}
\pubyear{1994}
\begin{document}
\label{firstpage}
\maketitle
\begin{abstract}
This paper examines a new class of exact and self-consistent MHD solutions 
which describe steady and axisymmetric hydromagnetic outflows from the 
atmosphere of a magnetized and rotating central object with 
possibly an orbiting accretion disk. 
The plasma is driven against gravity by a thermal pressure gradient, as well 
as by magnetic rotator and radiative forces. At the Alfv\'enic and fast 
critical points the appropriate criticality conditions are applied.  
The outflow starts almost radially but after the Alfv\'en transition and 
before the fast critical surface is encountered the magnetic pinching force 
bends the poloidal streamlines into a cylindrical jet-type shape. 
The terminal speed, Alfv\'en number, cross-sectional area of the jet, as 
well as its final pressure and density obtain uniform values at large 
distances from the source. 
The goal of the study is to give an analytical discussion of the 
two-dimensional interplay of the thermal pressure gradient, 
gravitational, Lorentz and inertial forces in accelerating and collimating 
an MHD flow.  
A parametric study of the model is also given, as well as a brief sketch 
of its applicability to a self-consistent modelling of collimated 
outflows from various astrophysical objects.
{The analysed model succeeds to give for the first time an exact and 
self-consistent MHD solution for jet-type outflows extending from the 
stellar surface to infinity where it can be superfast, in agreement with  
the MHD causality principle.}
\end{abstract}


\section{Introduction}

Collimated outflows are ubiquitous in astrophysics and cosmic jets are 
observed in the radio, infrared, optical, UV and X-ray parts of the spectrum, 
from the ground and space, most recently via the Hubble Space Telescope.  
Classes of objects in association with which jets are observed include young 
stellar objects (Ray 1996), old mass losing stars and planetary nebulae 
(Livio 1997), black hole X-ray transients (Mirabel \& Rodriguez 1996), 
supersoft X-ray sources (Kahabka \& Trumper (1996), high-mass X-ray 
binaries, cataclysmic variables (Shahbaz et al 1997) and many AGN and  
quasars (Biretta 1996, Ferrari et al 1996). 
Despite their observed abundance however, several key questions on their 
acceleration and collimation among others, have not been resolved yet. 

The theoretical MHD modelling of jets is not a simple undertaking, 
basically due to the fact that the set of the MHD equations is highly 
nonlinear with singular or critical points appearing in their domain of 
solutions;  
these singularities - through which a physical solution inevitably will have 
to pass - are not known a {\it priori} but they are determined only 
simultaneously with the complete solution.  
The purpose of the present study is to construct systematically a 
self-consistent MHD model for astrophysical jets
from the stellar base to infinity
where the interplay 
of the various forces acting on the plasma and which are able to 
accelerate and collimate the outflow, are analytically examined.  
This modelling is an improvement over the very few existing models developed 
so far to the same goal. For example, 
it is fully 2-dimensional ({\it cf.} Parker 1958, Weber \& Davis 1967), 
it does not contain singularities along the symmetry axis and the outflow 
is not overfocused but extends to large distances (as {\it e.g.} in  
Blandford \& Payne 1982, Ostriker 1997), 
the equation of state is not constrained by the artificial polytropic 
assumption (as {\it e.g.} in Contopoulos \& Lovelace 1994, Heyvaerts \&
Norman 1989),    
the thermal pressure is meridionally anisotropic ({\it cf.}  
Sauty \& Tsinganos 1994),  
the shape of the jet is self-consistently determined and not {\it a priori} 
given (as {\it e.g.} in Cao \& Spruit 1994, Kudoh \& Shibata 1997, 
Trussoni et al 1997), 
there is a steady asymptotic state ({\it cf.} Uchida \& Shibata 1985, 
Ouyed \& Pudritz 1997a,b, Goodson et al 1997), etc. {Furthermore, a gap 
the present model aspires to fill in the existing literature is the availability 
of a self consistent MHD model for jet-type outflows wherein the jet speed 
is superfast at large distances from the base such that all perturbations 
are convected downstream to infinity and they do not destroy the steady 
state solution. }

In the following Sec. 2 the basic steps for a systematic construction of 
this class of models are outlined. Then in Sec. 3 we discuss the critical 
surfaces which determine a physically interesting solution and in Sec. 4 the 
asymptotic behaviour of such solutions. 
A detailed parametric study of the model, including the 
solution topologies, is given in Sec. 5 and in the last Sec. 6 the connection 
of the dimensionless parameters characterizing the present model to the 
observable physical quantities of collimated outflows is briefly sketched.

\section{Construction of the model} 

In this section we describe in some detail how our model can be 
systematically obtained from the closed set of the governing MHD equations. 

\subsection{Governing equations}

The {\it kinematics} of astrophysical outflows may be described to zeroth 
order by the well known set of the steady (\(\partial/\partial t=0\))   
ideal hydromagnetic equations: 

\begin{equation}\label{mhdforce}
\rho \left( \vec{V}\cdot\vec{\nabla}\right)\vec{V}=
%\frac{1}{4\pi} \left(\vec{\nabla}\times
%\vec{B}\right)\times\vec{B}
\frac{\left(\vec{\nabla}\times
\vec{B}\right)\times\vec{B} } {4\pi}
-\vec{\nabla}P-\rho\vec{\nabla}{\cal V} +\vec{F}_{rad}
\,,
\end{equation}
\begin{equation}\label{mhdfluxes}
\vec{\nabla}\cdot\vec{B}=0\,,\quad 
\vec{\nabla}\cdot\left(\rho\vec{V}\right)=0
\,,\quad \vec{\nabla}\times\left(\vec{V}\times\vec{B}\right)=0
\,,
\end{equation}

\noindent
where \(\vec{B}\), \(\vec{V}\), \(-\vec{\nabla}{\cal V}=
-\vec{\nabla} \left( -{ {\cal GM}}/ {r} \right)\) denote the magnetic,
velocity and external gravity fields, respectively, $\vec{F}_{rad}$ the 
volumetric force of radiation while \( \rho\) and $P$ the gas density and 
pressure.\\
The {\it energetics} of the outflow on the other hand is governed by  
the first law of thermodynamics : 
\begin{equation}\label{firstlaw}
q=\rho \vec{V} \cdot \left [ \vec{\nabla}\left(\frac{1}{\Gamma-1}
\frac{P}{\rho}\right)+P \vec{\nabla} \frac{1}{\rho} \right ]
%=\rho \vec{V} \cdot \left [ \vec{\nabla}\left(\frac{\Gamma}{\Gamma-1}
%\frac{P}{\rho}\right)-\frac{1}{\rho}\vec{\nabla} P \right ]
\,,
\end{equation}
where $q$ is the volumetric rate of net energy input/output 
(Low \& Tsinganos 1986), while 
$\Gamma=c_{p}/c_{v}$ with $c_{p}$ and $c_{v}$ the specific heats for an 
ideal gas.\\
With axisymmetry in spherical coordinates $(r,\theta,\phi)$, the 
azimuthal angle $\phi$ is ignorable (\(\partial/\partial \phi=0\)) and   
we may introduce the poloidal magnetic flux function $A (r, \theta )$, such 
that three free integrals of $A$ exist. They are the total specific
angular momentum carried by the flow and magnetic field, $L(A)$, 
the corotation angular velocity of each streamline at the base of the flow,
$\Omega(A)$ and the ratio of the mass and magnetic fluxes, $\Psi_A(A)$
(Tsinganos 1982).
Then, the system of Eqs. (\ref{mhdforce}) - (\ref{mhdfluxes}) 
reduces to a set of two partial and nonlinear differential equations, 
i.e., the r- and $\theta$- components of the momentum equation on the 
poloidal plane. Note that by using the projection of the momentum equation 
along a stream-field line $A=const$ on the poloidal plane $(r,\theta)$, 
Eq. ({\ref{firstlaw}}) becomes, 
 
\begin{equation}\label{bernoulli}
\rho \vec{V}\cdot\vec{\nabla} \left(\frac{1}{2} V^2 + 
\frac{\Gamma}{\Gamma-1} \frac{P}{\rho}
+{\cal V} - 
\frac{\Omega r \sin \theta}{\Psi_A} B_{\phi} \right) -\vec{V}\cdot
\vec{F}_{rad}=q
\,.
\end{equation}

For a given set of the integrals $L(A)$, $\Omega (A)$ and $\Psi (A)$, 
equations (\ref{mhdforce}) - (\ref{mhdfluxes}) - (\ref{firstlaw})  
could be solved to give $\rho(r, \theta ) $, $P(r, \theta )$ and 
$A(r, \theta )$, \underbar{if} the heating function $q (r, \theta )$ 
and the radiation force $\vec{F}_{rad}$ are known. 
Similarly, one may close this system of Eqs.   
(\ref{mhdforce}) - (\ref{mhdfluxes}) - (\ref{firstlaw}), \underbar{if} an 
extra functional relation of $q$ with the unknowns $\rho$, $P$ and $A$ exists. 
As an example, consider the following special functional relation of $q$ 
with the unknowns $\rho$, $P$ and $A$ (Tsinganos et al 1992),
\begin{equation}\label{polytropic_heating}
q=\frac{\gamma-\Gamma}{\Gamma-1}
\frac{P}{\rho}\vec{V} \cdot \vec{\nabla}\rho
\,,
\end{equation}
where $\gamma \le \Gamma$. 
Then, Eq. (\ref{firstlaw}) can be integrated at once to give the familiar 
polytropic relation between $P$ and $\rho$, 
\begin{equation}\label{polytropic}
P=Q \left( A \right)\rho ^{\gamma}
\,,
\end{equation}
for some function $Q (A)$ corresponding to the enthalpy along a poloidal 
surface $A=const$. 
In this special case we can integrate the projection of the momentum 
equation along a stream-field line $A=const$ on the poloidal plane, 
Eq. (\ref{bernoulli}) by further assuming that 
$\vec{V}\cdot\vec{F}_{rad} = 0$, to get the well known Bernoulli integral
which subsequently can be combined with the component of the momentum 
equation across the poloidal fieldlines (the transfield equation) to yield 
$\rho$ and $A$.
After finding a solution, one may go back to Eq. (\ref{firstlaw}) and fully 
determine the function $q(r, \theta )$. 
It is evident that even in this special polytropic case 
with $\gamma \neq \Gamma$ the heating function $q$ (not its functional form 
but the function $q(r, \theta )$ itself) can be found only {\it a posteriori}.
Note that for $\gamma=\Gamma$ and only then the flow is isentropic. 
\\
Evidently, it is not possible to integrate Eq. (\ref{firstlaw}) for {\it any}  
functional form of the heating function $q$, such as it was possible with 
the special form of the heating function given in 
Eq. (\ref{polytropic_heating}).
To proceed further then and find other more general solutions (effectively 
having a variable value for $\gamma$), 
one may choose some other functional form for the heating function $q$ and 
from energy conservation, Eq. (\ref{firstlaw}), derive a functional form 
for the pressure.
Equivalently, one may choose a functional form for the pressure $P$ and 
determine the volumetric rate of thermal energy {\it a posteriori} 
from Eq. (\ref{firstlaw}), after finding the expressions of 
$\rho$, $P$ and $A$ which satisfy the two remaining components of the 
momentum equation. 
Hence, in such a treatment the heating sources which produce some specific 
solution are not known {\it a priori}; instead, they can be determined 
only {\it a posteriori}.
However, it is worth to keep in mind that as explained before, this situation is 
analoguous to the more familiar constant $\gamma$ polytropic case, with $\gamma 
\neq \Gamma $. In this paper we shall follow this approach, which is further 
illustrated in the following section.  

However, even with this approach, the integration of the system of mixed 
elliptic/hyperbolic partial differential equations (\ref{mhdforce}) - 
(\ref{mhdfluxes}) is not a trivial undertaking. Besides its nonlinearity, 
this is largely due to the fact that a physically interesting solution 
is bound to cross some critical surfaces which are not known {\it a priori} 
but they are determined simultaneously with the solution. For this reason 
only a very few such self-consistent solutions are available, albeit no one is 
superfast at infinity. 
Further assumptions on the shape of the critical surfaces are needed, as 
discussed in the following.

\subsection{Assumptions}

In order to construct analytically a new class of exact solutions, we shall 
proceed by making the following two key assumptions:
 \begin{enumerate}
  \item that the Alfv\'en number $M_{}^{}$ is some function of the
dimensionless radial distance $R={r/ r_{\star}}$, i.e., $M_{}^{} = M(R)$ \\
and
  \item that the poloidal velocity and magnetic fields have a dipolar
angular dependence,
\begin{equation}\label{assumptions2}
A= {r_{\star}^2 B_{\star} \over 2}{\cal A}\left(\alpha\right)\,,
\qquad
\alpha=\frac{R^2}{G^2\left(R\right)}\sin^2 \theta
\,,
\end{equation}
\end{enumerate}

\noindent
for some function $G(R)$. 

{A few words on the physical basis of the above two assumptions are needed at 
this point. These assumptions are expected to be physically reasonable for 
describing the outflow properties close to the rotation and magnetic axis 
of the system.  There, far from the distortion caused by the presence of an 
accretion disk, the {\it angular} distribution of the stellar magnetic 
field may be approximately dipolar (see Fig. 1 in Ghosh \& Lamb 1979). 
On the other hand, regarding the assumption of spherical critical surfaces, 
we note that all available numerical models derive a shape of 
the Alfv\'en surface which is approximately spherical near the rotation and 
magnetic axis of the system 
(Sakurai 1985, Bogovalov \& Tsinganos 1999, Ustyugova et al 1999).
Hence, although the analysed model may in principle extend to all angles 
from the symmetry axis since it satisfies the governing MHD equations, its 
physical applicability could be constrained around the polar regions only.}

We are interested in transAlfvenic flows and denote by a $\star$ the 
respective value of all quantities at the Alfv\'en surface. 
By choosing the function $G(R)$ such that $G\left(R=1\right)=1$ at 
the Alfv\'en transition $R=1$, it is evident that 
$G(R)$ measures the cylindrical distance $\varpi$ to the  
polar axis of each fieldline labeled by $\alpha$, normalized to its 
cylindrical distance $\varpi_{\alpha}$ at the Alfv\'en point, 
$G\left(R\right)={\varpi}/{\varpi_{\alpha}}$. For a smooth crossing of 
the Alfv\'en sphere $R=1$ [$r=r_{\star}, \theta = \theta_a (\alpha) $], 
the free integrals $L$ and $\Omega$ are related by 
\begin{equation}\label{regularity_phi}
{L\over \Omega } = 
\varpi_{\alpha}^2 (A) = r_{\star}^2 \sin^2 \theta_a (\alpha) =r_{\star}^2 
\alpha \,.
\end{equation}
Therefore, the second assumption is equivalent with the statement that  
at the Alfv\'en surface the cylindrical distance $\varpi_{\alpha}$ of each 
magnetic flux surface $\alpha=const$ is simply proportional to 
$\sqrt{\alpha}$. 
\\
%Note also that the gravitational potential can be expressed in 
%terms of the escape speed $V_{esc}$ at the Alfv\'en radius $r_\star$, 
%\[%\begin{equation}\label{gravity} 
%{\cal V}  =-\frac{\nu ^{2} V_{\star}^{2}}{2R}\,, \qquad 
%\nu=\frac{V_{esc}}{V_{\star}}\,,\qquad 
%V_{esc}=\sqrt{\frac{2 {\cal GM} }{r_{\star}}}\,.
%\]%\end{equation}
Instead of using the three remaining free functions of $\alpha$, 
(${\cal A}\,, 
\Psi_A$\,, $\Omega$), we found it more convenient to work instead with the 
three dimensionless functions of $\alpha$, ($g_1$\,, $g_2$\,, $g_3$),  
\begin{equation}\label{g1}
g_1\left( \alpha \right)= \int {\cal A}^{'2} d\alpha\,,
\end{equation}
\begin{equation}\label{g2}
g_2 \left( \alpha \right)=\frac{r_{\star}^2}{B_{\star}^2} \int \Omega ^2 \Psi_A ^2 d \alpha
\end{equation}
\begin{equation}\label{g3}
g_3\left(\alpha \right)=\frac{\Psi_A ^2 }{4\pi \rho_{\star}}\,. 
\end{equation}

These functions $g_{1}(\alpha),\,g_{2}(\alpha),\,
g_{3}(\alpha)$ are vectors in a 3D $\alpha$-space with some basis vectors  
$ u_1(\alpha)$, $u_2(\alpha)$,  $u_3(\alpha)$ 
(Vlahakis \& Tsinganos (1998), hereafter VT98).
Note that the forms of $g_{1} \,, g_{2} \,,g_{3}$ 
or equivalently the forms of $A$, $\Psi_{A}$, $\Omega $,
$L=r_{\star}^2 \alpha \Omega$ and $P$ should be such that the two remaining
components of the momentum equation are separable in
the $\alpha$ and $R$ coordinates.

\subsection{The method}

The main steps of the general method for getting exact solutions under the 
previous two assumptions are briefly the following. 

{\it First}, 
by using $\alpha$ instead of $\theta$ as the independent variable, 
we transform from the pair of the independent variables ($R\,, \theta$) 
to the pair of the independent variables ($R\,, \alpha$). 
The resulting form of the $\alpha$- component of the momentum 
equation can be integrated at once to yield for the gas pressure,
\begin{eqnarray}\label{pre}
\nonumber
P (R, \alpha )& =&\frac{B_{\star}^2}{8 \pi}
\left(f_0+f_4 g_{1}+ f_1 g_{1}^{'} + f_2 \alpha g_{1}^{'}+ f_5 g_{2}+  
f_3 \alpha g_{2}^{'} \right) 
\\ 
&=& 
\frac{B_{\star}^2}{8 \pi} {\bf{Y}}  {\bf{P}}^{\dag}
\,,
\end{eqnarray}
where $f_0\left(R\right)$ is a free function emerging from this integration,
$f_i\left(R\right)$, $i=1,2, \dots ,$  are functions of the spherical 
radius $R$ given in the 
Appendix  A and {\bf P} and {\bf Y} are the (1 $\times$ 7) matrices, 

\begin{equation}\label{Yinew}
{\bf{Y}}=
%\left[\;Y_1\,\;Y_2\,\;Y_3\,\;Y_4\,\;Y_5\,\;Y_6\,\;Y_7\;\right]=
\left[\; 1\,\;g_1\,\;g_1^{'}\,\;\alpha g_{1}^{'}\,\;g_2\,\;\alpha g_{2}^{'}\,
\;g_3\,\right]
\,,
\end{equation}

\begin{equation}\label{prenew}
{\bf{P}}=
%\left[\;p_1\,\;p_2\,\;p_3\,\;p_4\,\;p_5\,\;p_6\,\;p_7\;\right]=
\left[ \; f_0\, \; f_4\, \; f_1\, \; f_2\, \; f_5\, \; f_3\, \; 
0 \, \right]\,.
\end{equation}

%\begin{eqnarray}\label{Yinew}
%\begin{array}{l}
%{\bf{Y}}=\left[\;Y_1\,\;Y_2\,\;Y_3\,\;Y_4\,\;Y_5\,\;Y_6\,\;Y_7\;\right]=
%\\ 
%\left[\; 1\,\;g_1\,\;g_1^{'}\,\;\alpha g_{1}^{'}\,\;g_2\,\;\alpha g_{2}^{'}\,
%\;g_3\,\right]
%\,.
%\end{array}
%\end{eqnarray}
{\it Second}, by substituting Eq. (\ref{pre}) in the r-component of the 
momentum equation we obtain in terms of the (1 $\times$ 7) matrix  {\bf X}
\begin{equation}\label{Xinew}
{\bf{X}}=
%\left[\;X_1\,\;X_2\,\;X_3\,\;X_4\,\;X_5\,\;X_6\,\;X_7\;\right]=
\left[\;f_0^{'}\,\;f_4^{'}\,\;-\!f_6\,\;-\!f_7\,\;f_5^{'}\,\;-\!f_8\,\;-\!f_9\; 
\right] 
\,,
\end{equation}
%\begin{eqnarray}\label{Xinew}
%{\bf{X}}=\left[\;X_1\,\;X_2\,\;X_3\,\;X_4\,\;X_5\,\;X_6\,\;X_7\;\right]=
%\nonumber\\
%\left[\;f_0^{'}\,\;f_4^{'}\,\;-f_6\,\;-f_7\,\;f_5^{'}\,\;-f_8\,\;-f_9\; \right] 
%\,.
%\end{eqnarray}
the following equation:
\begin{eqnarray}\label{7-1}
%\mbox{or} \,, 
{\bf{Y}} {\bf{X}}^{\dag} = {\bf{0}}
\,. 
\end{eqnarray}

%Note that a 
A key step in the method is to find a possible set of
vectors $ u_1(\alpha)$,  $u_2(\alpha)$, $u_3(\alpha)$ such that all 
components of the matrix {\bf Y} belong to the same $\alpha$-space.   
To that goal we choose $ u_1(\alpha)=1$ and $u_2(\alpha)=g_1(\alpha)$.
If this is the case, then our  {\it third} step is  
to construct a $3 \times 7$ matrix ${\bf {K}}$ such that 
\begin{equation}\label{yk}
{\bf{Y}}=\left[\;{ u_1}\,\;{ u_2}\,\;{u_3}\;\right] 
{\bf{K}}
\,.
\end{equation}
Then, from  Eq. (\ref{7-1}),
\[
\left[\;{ u_1}\,\;{u_2}\,\;{ u_3}\;
\right]\,{\bf {K}} {\bf{X}}^{\dag}={\bf{0}} \,,
\] 
and since ${u_i}$ are linearly independent it follows  
\begin{equation}\label{kx}
{\bf{K}} {\bf{X}}^{\dag}={\bf{0}} 
\,.
\end{equation}

{\it Finally}, it follows from Eq. (\ref{pre}) , (\ref{prenew}) and 
(\ref{yk}) that,
\[
P
%=\frac{B_{\star}^2}{8 \pi} \left[\;{ u_1}\,\;{ u_2}\,\;{ u_3}
%\;\right] {\bf{K}} {\bf{P}} ^{\dag}
= \frac{B_{\star}^2}{8 \pi} \left( P_0 + g_1 P_1+ u_3 P_2
\right)
\,, \]
where the three components of the pressure $P_0$, $P_1$ and $P_2$ are
\begin{equation}\label{prec}
[P_0 \; P_1 \; P_2]^{\dag}={\bf K} {\bf P}^{\dag}
\,.
\end{equation}

\subsection{The Model}

Let us know apply this method in the construction of a specific model.   
We may recall that in a previous paper, (VT98), 
it was found that only nine distinct general families of such 
vectors exist. One of them is, 

\begin{equation}\label{unitv}
u_1(\alpha)=1\,,\quad  u_2(\alpha)=\alpha\,,\quad  
u_3(\alpha)=\alpha^{\epsilon}
\,, 
\end{equation}
while the corresponding free functions are, 

\begin{equation}
g_1(\alpha ) = a 
\,,
\end{equation}
\begin{equation}
g_2(\alpha ) = \xi \alpha+{\mu \alpha^{\epsilon}/\epsilon}
\,,
\end{equation}
\begin{equation}
g_3(\alpha ) = 1+\delta \alpha +\mu \delta_{0} \alpha^{\epsilon}
\,.
\end{equation}
%for $\epsilon \neq 0,1 ,\mu \neq 0$. 

\noindent
For these particular choice of $u_1(\alpha)=1\,, u_2(\alpha)=\alpha\,,  
u_3(\alpha)=\alpha^{\epsilon}$ we find the following form of the matrix
{\bf K}, 
\begin{equation}\label{k}
{\bf{K}}=
\left[
\begin{array}{ccccccc}
1 & 0 & 1 & 0 & 0 & 0 & 1 \\
0 & 1 & 0 & 1 & \xi & \xi & \delta \\
0 & 0 & 0 & 0 & \displaystyle \frac{\mu}{\epsilon} & \mu & \mu \delta_0 
\end{array}
\right] \,.
\end{equation}

Then, from Eqs. (\ref{prenew}) and (\ref{prec}) we get,

\begin{equation}\label{P123}
\left[\begin{array}{c}
P_0 \\ P_1 \\ P_2 
\end{array}
\right] =\left[
\begin{array}{l}
f_0+f_1 \\ f_4+f_2+\xi \left(f_3+f_5 \right) \\
\mu \left(\displaystyle \frac{f_5}{\epsilon}+f_3 \right)
\end{array}
\right]\,.
\end{equation}
 
Finally, from Eq. (\ref{kx}) using the definitions of  
Eqs. (\ref{Xinew}), (\ref{k})  we obtain,
three \underline{ordinary} differential equations, the system (\ref{arxdiaf}), 
for the functions of $R$ in the model 
for $\epsilon \neq 0,1$ and $\mu \neq 0 $ 
(only then we have a 3D $\alpha$-space with $1 \,, \alpha \,, 
\alpha^{\epsilon}$ linearly independent).
%figure 1
\begin{figure}
\centerline{
\psfig{file=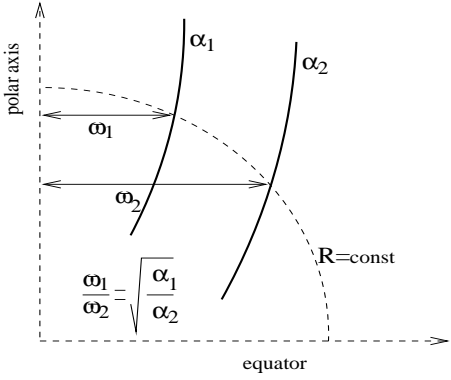,height=5.0truecm,width=6.0truecm,angle=360}}
\caption{\label{sketch11}
An illustration of the construction of the 
streamlines $\alpha=const.$ 
on the poloidal plane in meridionally selfsimilar outflows.}
\end{figure}
Altogether, let us summarize the characteristics of our model.
The physical quantities of the outflow have the following exact expressions:
\begin{equation}\label{density}
\rho=\frac{\rho_{\star}}{M^2} \left(1+\delta \alpha+\mu \delta_0 \alpha^{\epsilon}\right)
\end{equation}
\begin{equation}\label{pressure}
P=\frac{\rho_{\star} V_{\star}^{2}}{2}
\left(P_{0}+P_{1}\alpha +P_{2}\alpha^{\epsilon}\right),
\end{equation}
\begin{eqnarray}\label{V}
\nonumber
\vec{V}=V_{\star}\left(\frac{M^2\cos{\theta}}{G^2}\hat{r}-\frac{M^2 F \sin{\theta}}
{2G^2}\hat{\theta}+
\right.
\\
\left.
 \sqrt{\xi \alpha+\mu \alpha^{\epsilon}}\frac{G^2-M^2}
{G\left(1-M^2 \right)}\hat{\phi}\right)/\sqrt{1+\delta \alpha+\mu \delta_0 \alpha^{\epsilon}}
\end{eqnarray}
\begin{eqnarray}\label{B}
\nonumber
\vec{B}=B_{\star}\left(\frac{\cos{\theta}}{G^2}\hat{r}-
\frac{F\sin{\theta}}{2G^2}\hat{\theta}-
\right.
\\
\left.
\sqrt{\xi \alpha+\mu\alpha^{\epsilon}}
\frac{1-G^2}{G\left(1-M^2 \right)}
\hat{\phi}\right)
\,, %I \propto \varpi B_{\phi} \propto \frac{1-G^2}{1-M^2}\sqrt{\xi \alpha^2+ \mu \alpha^{\epsilon+1}}
\end{eqnarray}

\noindent
where the five unknown functions $G^2(R)$, $F(R)$, $M^2(R)$, $P_1(R)$ and 
$P_0(R)$ entering in the above expressions are obtained from the 
integration of the
five first order ordinary differential equations given in Appendix B,  
while
the pressure component $P_2(R)$ is given explicitly in terms of 
the other variables (Appendix B).

{Note that the parameters $\epsilon$ and $\xi$ determine the enclosed 
poloidal current by a given magnetic surface in the outflow.
The parameters $\epsilon\,,\delta\,,\delta_0$ determine the latitutianal
distribution of the density. The parameter $\mu$ controls the differential
rotation. Further discussion of the physical meaning of these parameters
will be given later (Secs. 2.5, 5).}

\subsection{Some properties of the meridionally self-similar model}

Our model is meridionally self-similar, i.e., if we know the shape of 
one fieldline $\alpha = \alpha_1$ we may derive the shape of any other  
streamline $\alpha = \alpha_2$ by moving in the meridional direction 
along each cycle $R=const$ on the poloidal plane as illustrated in Fig. 
(\ref{sketch11}).
%${\varpi_1}/{\varpi_2}=\sqrt{{\alpha_1}/{\alpha_2}}$.

Note that the flux function $A$ is simply proportional to $\alpha$ 
which means that for cylindrical solutions at $R \gg 1 $, 
the magnetic field on the poloidal plane is uniform and its strength  
is independent of $\alpha$, 
$\mid \vec{B_p}\mid _{\infty}  ={B_{\star}}/{G_{\infty}^2}$.\\
%In this regime all forces caused from the rotation.\\

\begin{figure}
\centerline{
\psfig{file=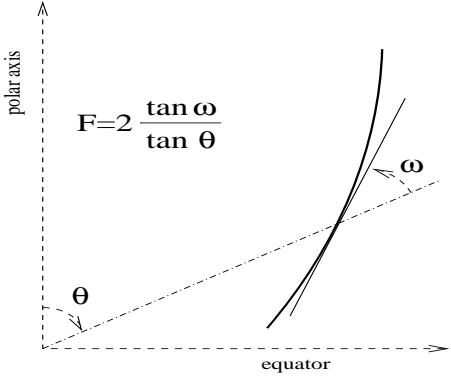,height=5.0truecm,width=6.0truecm,angle=360}}
\caption{\label{sketch2}
A geometrical illustration of the expasion factor $F(R)$ 
which determines the shape of the poloidal streamlines in a meridionally 
selfsimilar outflow.}
\end{figure}

The density at the Alfv\'en surface is 
$$
\rho_{\alpha}=\frac{\Psi_{A}^{2}}{4 \pi }=\rho_{\star}
\left(1+\delta \alpha+\mu \delta_0 \alpha^{\epsilon}\right)
\,,
$$ 
i.e., it is similar to a Taylor expansion in the cylindrical distance 
$\varpi_{\alpha}$ from the rotation and magnetic axis $\alpha = 0$. 
For example, for $\epsilon=0.5$ we have, 
$$
\frac{\rho_{\alpha}}{\rho_{\star}}=
1+\mu \delta_0 \frac{\varpi_{\alpha}}{r_{\star}}+
\delta \left( \frac{\varpi_{\alpha}}{r_{\star}} \right)^2
\,.
$$
We 've also introduced the expansion factor 
$$
F\equiv  \frac{\partial \ln \alpha (R, \theta) }{\partial \ln R}=
2-R\frac{G^{2'}}{G^2}
\,,
$$
which measures the opening of the fieldlines on the poloidal plane, 
as illustrated in Fig. (\ref{sketch2}).
%figure 2 
Thus, if  $F>2$ the fieldlines turn towards the axis, if $F=2$ 
they expand cylindrically, if $F=0$ they are purely radial while  
if $F<0$ the fieldlines turn toward the equator (in that case, there is 
a closed region near the equator).
If we eliminate $F$ in Eq. (\ref{dfdr}) (using Eq. (\ref{dg2dr}) )
we have the second derivative of $G$ (which corresponds to the term 
$\frac{\partial^2 A}{\partial r^2}$ of the transfield equation).
So, using $F$ us an intermediate function we have only first order 
differential equations.

\subsection{Radiative acceleration}

For the radiative acceleration we have assumed two components. 
The first component is due to continuum absorption and is set proportional 
to the radiative flux. It drops with distance as $r^{-2}$, similarly 
to gravity. If $L_{\varepsilon}$ is the Eddington luminocity,  
we may use the ratio $\Gamma_{\varepsilon}={L}/{L_{\varepsilon}}$
such that this part of the radiative acceleration is 
$\Gamma_{\varepsilon} \rho {{\cal GM}}/{r^2}$.\\
We have also assumed a second component of the radiative acceleration 
due to line contribution. By adopting the optically thin atmosphere 
aproximation (Lamers, 1986, Chen \& Marlborough 1994, Kakouris \& Moussas 1997),
this part of the acceleration is simply a function of $r$ 
since in general the total number of weak lines is a function of $r$.
Then, the corresponding expression of the radiative acceleration is 
${V_{\star}^2 }/{r_{\star}} \rho Q\left(R\right)$.

The combination of gravitational and radiative acceleration is thus 
$$
-\rho \vec{\nabla}{\cal V} +\vec{F}_{rad}=
\frac{V_{\star}^2 }{r_{\star}} \rho \left( Q \left(R\right)-\frac{\nu^2}
{2 R^2} \right) \hat{r}
\,,
$$ 
where 
$$\nu^2=\frac{V_{esc}^2}{V_{\star}^2}\left(1-\Gamma_{\varepsilon}\right)=
\frac{2 {\cal GM}}{r_{\star} V_{\star}^2 }\left(
1-\Gamma_{\varepsilon}\right)
\,.
$$
%The function $f_9$ is now 
%\begin{equation}\label{newf9}
%f_9=\frac{2}{M^2} \left(Q-\frac{\nu^2}{2R^2}\right) \,.
%\end{equation}
Furthermore, we use for $Q$ the aproximation  of a power law, 
$Q \left(R\right)={\mu_0}/{R^x}$ with $\mu_0$, and $x$ constants.
\\
%If we substitute the forms of $\rho,P,{\cal V},\vec{V},\vec{B}$ 
%and $\vec{F}_{rad}$ in the remaining $ -\hat{r}$
%and $ -\hat{\theta}$ components of the momentum equation 
%we find the ordinary 
%differential equations for the unknown functions of $R$ , 
%the system (\ref{arxdiaf}).

In the following we shall discuss the results of the integration.
%of the
%previous system of differential equations (\ref{dg2dr}), (\ref{dfdr}), 
%(\ref{dm2dr}), (\ref{dp1dr}), (\ref{dp0dr}), (\ref{p2}).
Finally, we shall calculate all the other remaining physical quantities. 
%including the volumetric rate of thermal energy $q$.
A parametric study will be made only for $\epsilon > 0$, since for 
$\epsilon <0$ we have $\rho \rightarrow \infty $ as $\alpha \rightarrow 0$.\\

For $\epsilon =0$ or $1$ (or equivalently for $\mu=0$), we get a degenerate 
case which needs an extra condition between the functions of $R$. 
This case has been studied in Sauty \& Tsinganos (1994) (where the components 
of the pressure $P_0\,,P_1$ are set proportional to each other)
and Trussoni et al (1997) (where the function $G(R)$ is given {\it a priori}).
Here, in the case $\mu=0$ we 've chosen this extra condition to be 
${f_5^{'}}/{\epsilon}-f_8-\delta_0 f_9=0$
({\it c.f.}, the last equation of the system (\ref{arxdiaf}).
%\newpage
\section{Critical Surfaces}

In the domain of the solutions there exist several critical surfaces. In the 
following we briefly discuss the physical context of these critical 
surfaces. 

\subsection{Alfv\'en critical surface}

We recall that one of our goals is to investigate 
transalvfenic solutions wherein $L=\varpi_{\alpha}^2 \Omega$. 
By multiplying Eq. (\ref{dfdr}) with $1-M^2$ and evaluating the resulting 
expression at the Alfv\'en point we get \\
\begin{equation}\label{alfcrit}
F_{\star} p_{\star} -\frac{F_{\star}^2-4}{2} -2 P_{1\star} = 0\,,
\end{equation}
\vspace{0.3cm}
with $F_{\star}$, $P_{1\star}$ and 
$p_{\star}=\displaystyle{\left({dM^2}/{dR}\right)_{\star}}$ 
the respective values of these quantities at the Alfv\'en transition $R=1$.
Eq. (\ref{alfcrit}) is the so-called Alfv\'en regularity condition 
in the present model.
Note that if we also multiply Eq. (\ref{dp1dr}) with $1-M^2$ and evaluate 
the resulting expression at the Alfv\'en point we get an identical
expression while Eq. (\ref{dm2dr}) after using L'Hospital's rule gives 
an identity.

\subsection{Slow/fast critical surfaces.}

In order to locate the critical surfaces where the radial component 
of the flow speed equals to the corresponding slow/fast MHD wave speeds 
(Tsinganos et al 1996), we need to calculate first the sound speed 
$C_s$ at these points; to this goal we may proceed as follows.
\\   
Consider that at some fixed distance $R$ of a given streamline labeled by 
$\alpha$ we make a small change in the density $\rho$ and the pressure $P$. 
We may define the square of the sound speed as the ratio of such an 
infinitesimal change of $P$ and $\rho$,\\ 
\noindent
\begin{eqnarray}\label{cs2}
\displaystyle{
\nonumber
C_s^2=\left(\frac{\partial P}{\partial \rho}\right)_{\alpha,R}=
-\frac{V_{\star}^2}{2}
\frac{M^4
}
{\displaystyle{ 1+\delta \alpha +
\mu \delta_0 \alpha^{\epsilon}}}
}
\times
\\
\displaystyle{ \left(
\frac{\partial P_0\left(R,M^2\right)}{\partial M^2}+
\frac{\partial P_1\left(R,M^2\right)}{\partial M^2}\alpha +
\frac{\partial P_2\left(R,M^2\right)}{\partial M^2}\alpha^{\epsilon} \right)
}
\,,
\end{eqnarray} 
\vspace{0.5cm}
using Eqs. (\ref{density}) - (\ref{pressure}).
%\noindent
But from the differential equation (\ref{dp0dr}) we can calculate
the derivative ${\partial P_0\left(R,M^2\right)}/{\partial M^2}$ 
while from Eq. (\ref{dp1dr}) after substituting $dF/dR$ from Eq. (\ref{dfdr}) 
we can calculate the other derivative 
${\partial P_1\left(R,M^2\right)}/{\partial M^2}$.
%\begin{equation}\label{partp0}
%\displaystyle{
%\frac{\partial P_0\left(R,M^2\right)}{\partial M^2}=-\frac{2}{G^4}
%}
%\,,
%\end{equation} 
%\vspace{0.5cm}
%\begin{equation}\label{partp1}
%\displaystyle{
%\frac{\partial P_1\left(R,M^2\right)}{\partial M^2}=
%-\frac{F^2-4}{2 R^2 G^2} -
%2 \xi \frac{\left(1-G^2\right)^2}{G^2 \left(1-M^2\right)^3}-
%\frac{M^2 F^2}{2 R^2 G^2 \left(1-M^2\right)}
%}
%\,.
%\end{equation} 
%\vspace{0.3cm}
%\noindent
Finally, from Eq. (\ref{p2}) by taking the derivative of $P_2(G, M^2)$ 
for constant $G(R)$ we get similarly
the derivative ${\partial P_2\left(R,M^2\right)}/{\partial M^2}$.
%\begin{equation}\label{partp2}
%\displaystyle{
%\frac{\partial P_2\left(R,M^2\right)}{\partial M^2}=
%\frac{\mu}{G^2} \left[
%\frac{\left(2 M^2-1\right) G^4 -M^4}{\epsilon M^4 \left(1-M^2\right)^2}-
%2\frac{\left(1-G^2\right)^2}{\left(1-M^2\right)^3}\right]}
%\,.
%\end{equation}
%\vspace{0.4cm}
Substituting
%Eqs. (\ref{partp0}), (\ref{partp1}) and (\ref{partp2}) 
these derivatives
in Eq. (\ref{cs2}) we obtain the expression of the sound speed at
the points where $\left(2 M^2-1\right) G^4 -M^4=0$.
%.(Note that for $\alpha=0,C_s=V_r$.)
\\
An inspection of Eq. (\ref{dm2dr}) for the Alfv\'en number 
$M(R)$ reveals that besides the Alfv\'en transition where 
$M=G=R=1$, there may be other distances $R_x \neq 1$ where the 
denominator of this equation becomes zero, ${\cal D} \equiv 
\left[\left(2 M^2-1\right) G^4 -M^4\right]_{R=R_x}=0$. In such a case, the 
numerator of Eq. (\ref{dm2dr}) should be also set equal to zero and we 
have conditions typical of a critical point (using L'Hospital's rule 
we find two solutions for the slope of $M^2$ in this point, i.e., this 
singularity corresponds to an x-type critical point).
To clarify the physical 
identity of such a critical point, we may manipulate the denominator 
${\cal D}$ and write it in the form \\
\begin{eqnarray}\label{modify}
\nonumber
\displaystyle{
{\cal D} =2 \epsilon G^2 
\frac{1+\delta \alpha +\mu \delta_0 \alpha^{\epsilon}}
{V_{\star}^2 V_{ A,r}^4 \mu \alpha^{\epsilon}} } \times
\\ 
\displaystyle{\left(V_r^2-V_{ A, r}^2\right) 
\left[ V_r^4-V_r^2 \left(C_s^2+ V^2_{ A} \right)+
C_s^2 V^2_{ A, r} \right]
}
\,,
\end{eqnarray}
\vspace{0.3cm}
\noindent
where $V_{ A}$, $V_{ A, r}$ are the total and radial Alfv\'en 
speeds, respectively. 
Evidently, a critical point at $R_x$ corresponds to the modified by 
the meridional self-similarity fast/slow critical points (Tsinganos et al 
1996). In other words, the sphere $R=R_x$ is the corresponding 
spherical separatrix in the hyperbolic domain of the system of the 
MHD differential equations (Bogovalov 1996).
The sound speed is well defined at the critical points where ${\cal D}=0$,
but it is an open question if this definition can be extended everywhere.
%\newpage
\section{Asymptotic analysis}

According to the asymptotical behaviour of the poloidal streamlines we may 
distinguish two different types of solutions.

\subsection{Cylindrical asymptotics achieved through oscillations 
(Type I solutions)}

In this case the poloidal streamlines undergo oscillations of decaying 
amplitude and finally they become cylindrical. A similar oscillatory 
behaviour is found in all physical quantities, a situation which has been 
already analysed in detail (Vlahakis \& Tsinganos, 1997, henceforth VT97). 
According to this analysis, as $R \gg 1$ we have  
\begin{equation}
M_{}^{2}=M_{\infty}^{2}\left(1+\lambda_{0}\varepsilon\right)\,,
\quad
G^2=G_{\infty}^{2}\left(1-\varepsilon\right)
\,,
\end{equation}
\begin{equation}
\varepsilon (r) \approx \frac{D}{r^s} \sin{\left(kr+\phi_0\right)}\,,
\quad
s=2+\frac{\lambda_{0}M_{\infty}^{2}}{M_{\infty}^{2}-1}
\,,
\end{equation}
\begin{equation}\label{V4}
k^{2}=\frac{2\xi\left(1-\epsilon\right)\left(M_{\infty}^{2} -G_{\infty}^{4}
\right)}
{r_\star^{2}M_{\infty}^{2}\left(1-M_{\infty}^{2}\right)^{2}} 
\,,
\end{equation}
\begin{equation}\label{V5}
\lambda_{0}=\left[\left(\epsilon+1\right)M_{\infty}^{2}
-\left(\epsilon-1\right)G_{\infty}^{4}\right]\frac{1-M_{\infty}^{2}}{
\left(2 M_{\infty}^{2}-1\right)G_{\infty}^{4}-M_{\infty}^{4}}
\,.
\end{equation}
Note that for $s > 1$ the gravitational term is dominant, but the analysis 
is still correct because the oscillatory perturbation is independent of the 
"backround" term $1 / r$ (VT97).

\subsection{Converging to the axis asymptotics (Type II solutions)}

An analysis of the system of the differential equations 
(\ref{dfdr}) - (\ref{dp0dr}) for 
$G (R \rightarrow \infty ) \rightarrow 0$, $M (R \rightarrow \infty ) 
\rightarrow \infty$ and $F (R \rightarrow \infty ) \rightarrow F_{\infty}$ 
shows that 
in this case the value of the expansion factor $F_{\infty}$ at 
$R \gg 1$  approaches a constant value, the positive root of the equation \\
$$
\left(\epsilon+3\right)\left(\epsilon+\frac{3}{2}\right)F_{\infty}^2-
2 \left(\epsilon+2\right)\left(\epsilon+\frac{5}{2}\right)F_{\infty}-
4 \left(\epsilon+2\right) =0
\,.
$$

\vspace{0.3cm}
As we shall see later, solutions are obtained mainly for 
$\epsilon > 0$, in which case this root is greater than 2, $F_{\infty} > 2$, 
i.e., the cross-sectional area of flow tube drops to zero at large radii, 
$G^2 \propto R^{2-F_{\infty}}$. 
The poloidal velocity goes to 
infinity as $V_r \propto R^{\left(\epsilon+2\right)
\left(F_{\infty}-2\right)}$ to conserve mass,
while the toroidal velocity grows like  
$V_{\phi} (R \rightarrow \infty ) \propto R^{F_{\infty}-2}$ from 
angular momentum conservation.

\section{Parametric study of solutions}

The two crucial parameters which affect the qualitative behaviour of the 
model are $\xi$ and $\epsilon$.\\
{\it First} for $\epsilon$, from the expression of the density 
$\rho$ in Eq. (\ref{density}) it is required that $\epsilon>0$ in order 
that the density at the axis, $\rho(\alpha = 0 ,R)$ 
and the pressure are finite. In the case 
$\epsilon=1$ the electric current $I_z(\alpha ,R)$ enclosed by a poloidal 
magnetic flux tube $\alpha = const.$ and the corresponding confining 
azimuthal magnetic field $B_{\phi} (\alpha ,R)$ are proportional to 
$\alpha$; this case has been already studied in Sauty \& Tsinganos (1994)
and it was found that 
cylindrical asymptotics is obtained through oscillations. 
If $\epsilon>1$, $I_z(\alpha ,R)$ and  $B_{\phi } (\alpha ,R)$ are 
increasing faster with $\alpha$ which results in a stronger magnetic 
pinching force which eventually reduces the cross-sectional area of the 
flow tube to zero. Therefore we expect 
that when $0<\epsilon<1$ we obtain asymptotically cylindrical solutions 
while for larger values solutions where asymptotically $G_{\infty} 
\longrightarrow 0$,  as it may be seen in Figure (11).  
For the larger values of $\epsilon > 1$ the pinching is so strong that 
oscillations do not exist. This may also be seen from Eq. (\ref{V4}) where 
$k^2 <0$ for $\epsilon > 1$ (for $\xi (M^2_{\infty} -G_{\infty}^{4})>0$). 
Note that if $\epsilon >1$, it is needed to have $\xi >0$ such that the 
square roots in Eqs. (\ref{omega}), (\ref{L}) are positively defined near the axis 
$\alpha \approx 0$. 

Altogether then, we shall divide accordingly our parametric study to the 
intervals $0< \epsilon<1$,  for cylindrical asymptotics with oscillations  
[cases (a)-(b)] and $\epsilon >1$, for converging to the axis fieldlines 
without oscillations [case (c)] .\\
{\it Second},  the parameter $\xi$ is related to the asymptotic value of 
the pressure component $P_1$
(and through force balance in the cylindrical direction to $B_{\phi}$ and $I_z$). 
For cylindrical solutions at $R \gg 1$ we get from the asymptotic analysis \\
$$
P_{1,\infty}=
%\xi \left(\frac{M^4-\left(2 M^2-1\right)G^4-2 M^2 \left(1-G^2\right)}
%{G^2 M^2 \left(1-M^2\right)^2}\right)_{\infty}
%=
-\xi \left(\frac{
\left(M^2-1\right)\left(G^4-M^2\right)+M^2\left(1-G^2\right)^2
}{G^2 M^2 \left(1-M^2\right)^2}\right)_{\infty}
\,.
$$

For example when $\xi>0$, in which case from the integration we find 
$G_{\infty}<1 \ll M_{\infty}$, we obtain $P_{1,\infty} >0$ and 
the pressure gradient assists the magnetic pressure in collimating the
outflow. In that respect solutions with $\xi>0$ correspond to an  
underpressured jet (Trussoni et al 1997).
On the other hand when $\xi <0$ in which case from the integration 
$G_{\infty}^{4}>M^2_{\infty}>1$ and we find $P_{1,\infty} >0$, 
$P_{2,\infty} <0$.
In all solutions with cylindrical asymptotics (i.e., for $\epsilon<1$),
one finds that for $\xi>0$ the total pressure force in
the $\hat{\varpi}$ direction $-\hat{\varpi} \vec{\nabla}\left(P+\frac{B^2}{8 \pi}\right)$ is 
towards the axis while for $\xi<0$ it is in the opposite direction.
\setbox1=\vbox{\hsize=11truecm\vsize=10truecm
\psfig{figure=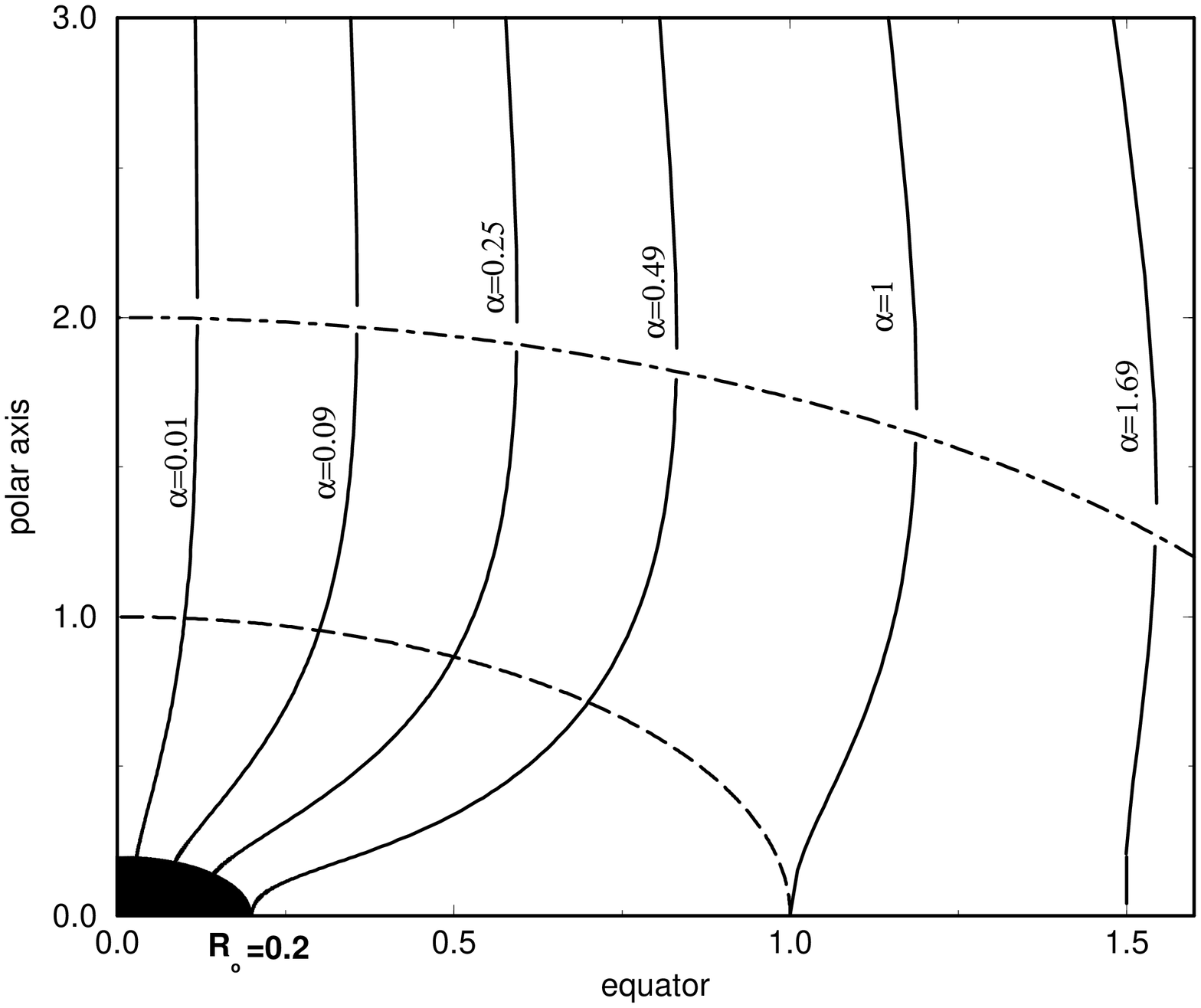,height=10.0truecm,width=8truecm,angle=270}}
\setbox2=\vbox{\hsize=11truecm\vsize=10truecm
\psfig{figure=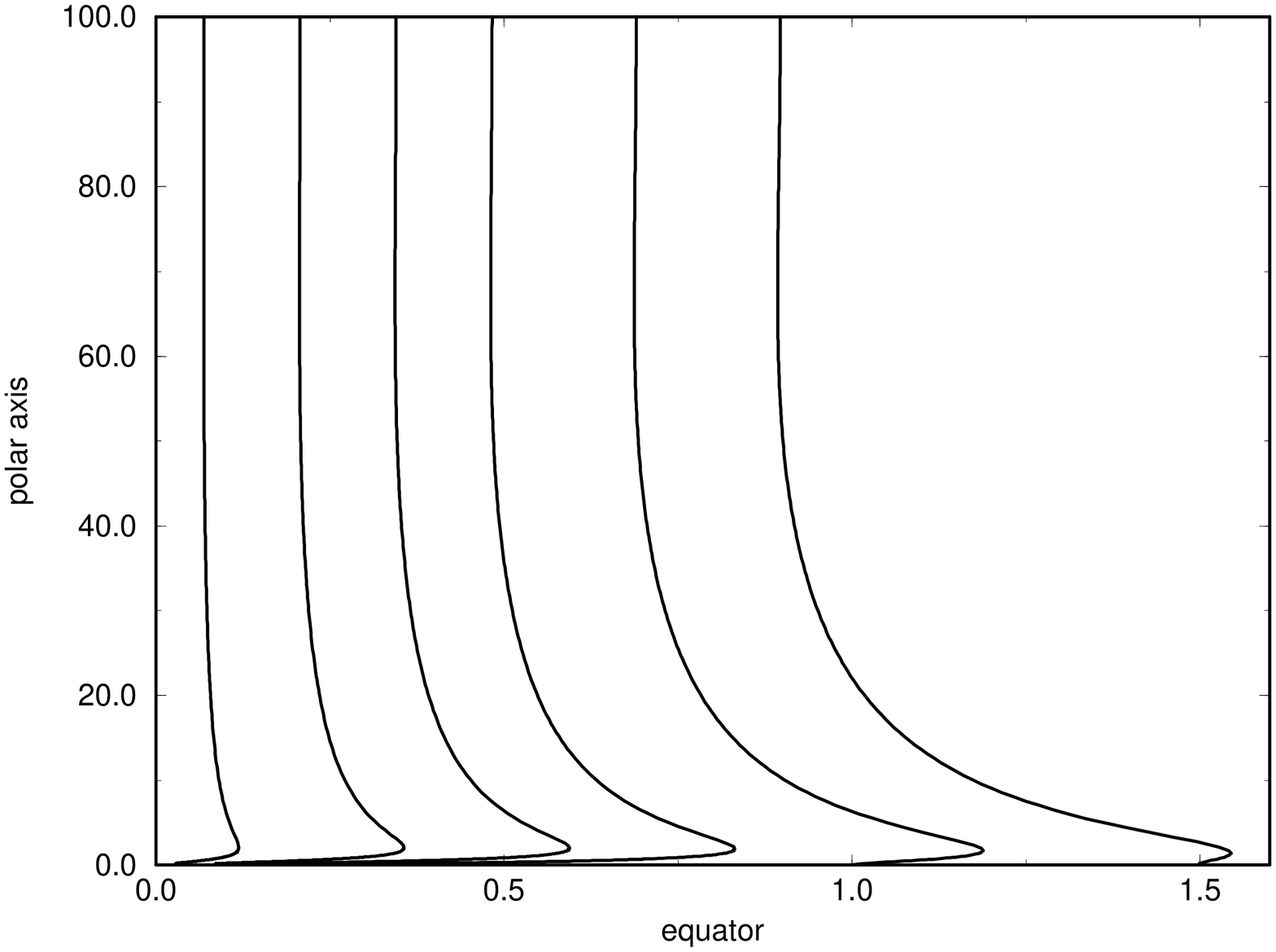,height=10truecm,width=8truecm,angle=270}}
%figure 3
\begin{figure*}
\centerline{
%\hspace{-1.5cm}
\box1
\hspace{-4.cm}
\box2}
\caption{\label{stream_a}
Poloidal streamlines close and far from the central object
for {\bf case (a)} with parameters: 
$\epsilon=0.5$, $\xi=10$, $\delta \nu^2=3.5$, $\delta_0 \nu^2=0.1$, 
$\mu_0=0$, $F_{\star}=1$ and $p_{\star} \approx 2.2655 $.
The Alfv\'en (fast) surface is indicated by dashed (dot-dashed) lines.}
\end{figure*}
%figure 4
\begin{figure}
\centerline{\psfig{file=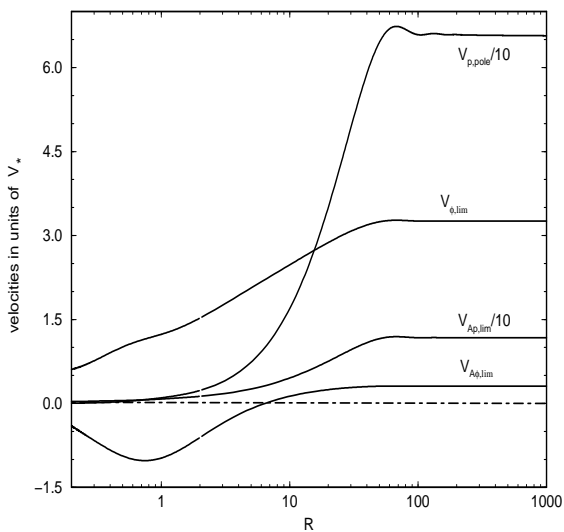,height=8.5truecm,width=8.5truecm,angle=270}}
\caption{\label{vel_a}
Dimensionless velocities for case {\bf (a)} with parameters: 
$\epsilon=0.5$, $\xi=10$, $\delta \nu^2=3.5$, $\delta_0 \nu^2=0.1$, 
$\mu_0=0$, $F_{\star}=1$ and $p_{\star} \approx 2.2655 $}
\end{figure}
\setbox1=\vbox{\hsize=8.0 truecm \vsize=8.0truecm
\psfig{figure=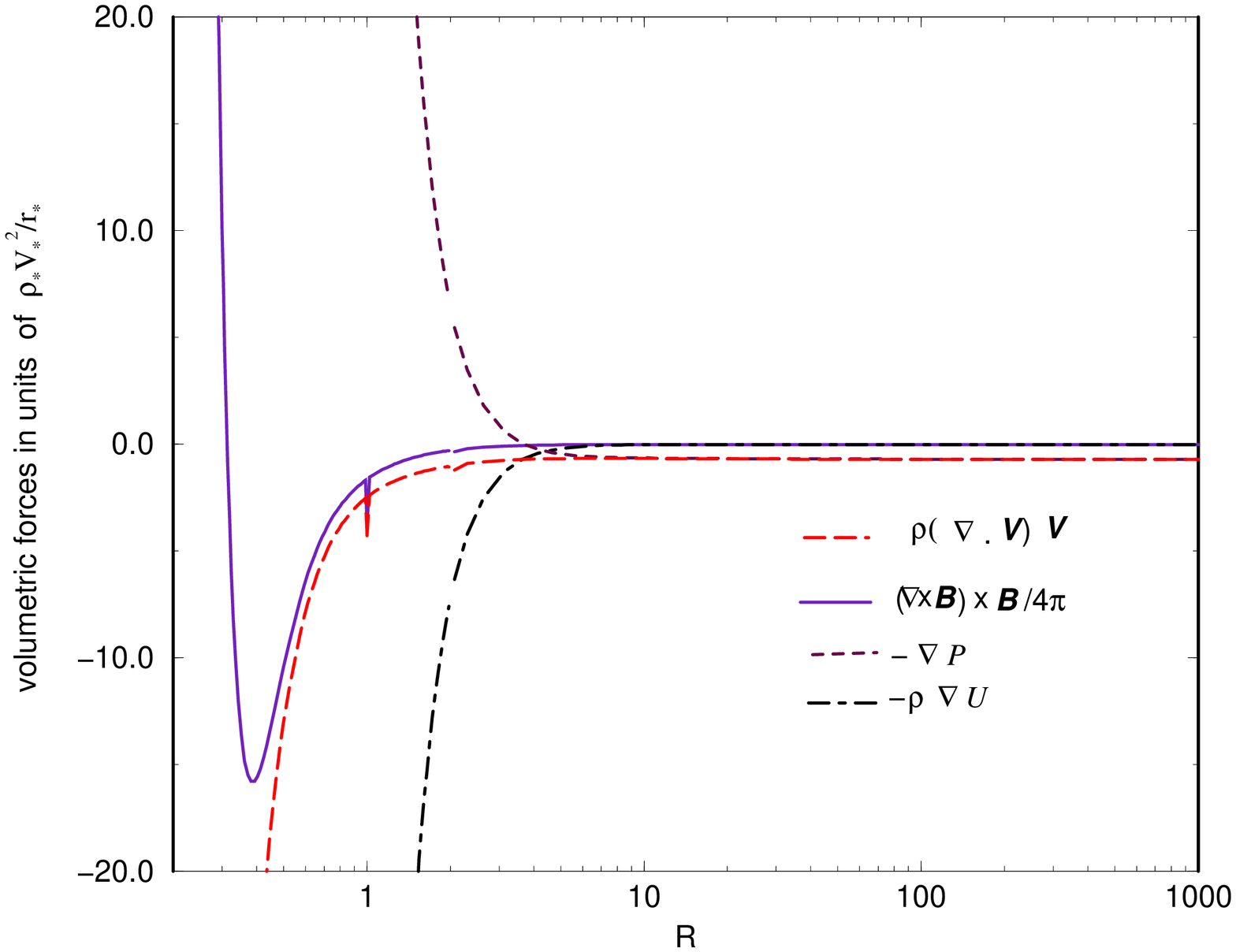,height=8.0truecm,width=8.0truecm,angle=270}}
\setbox2=\vbox{\hsize=8.0 truecm \vsize=8.0truecm
\psfig{figure=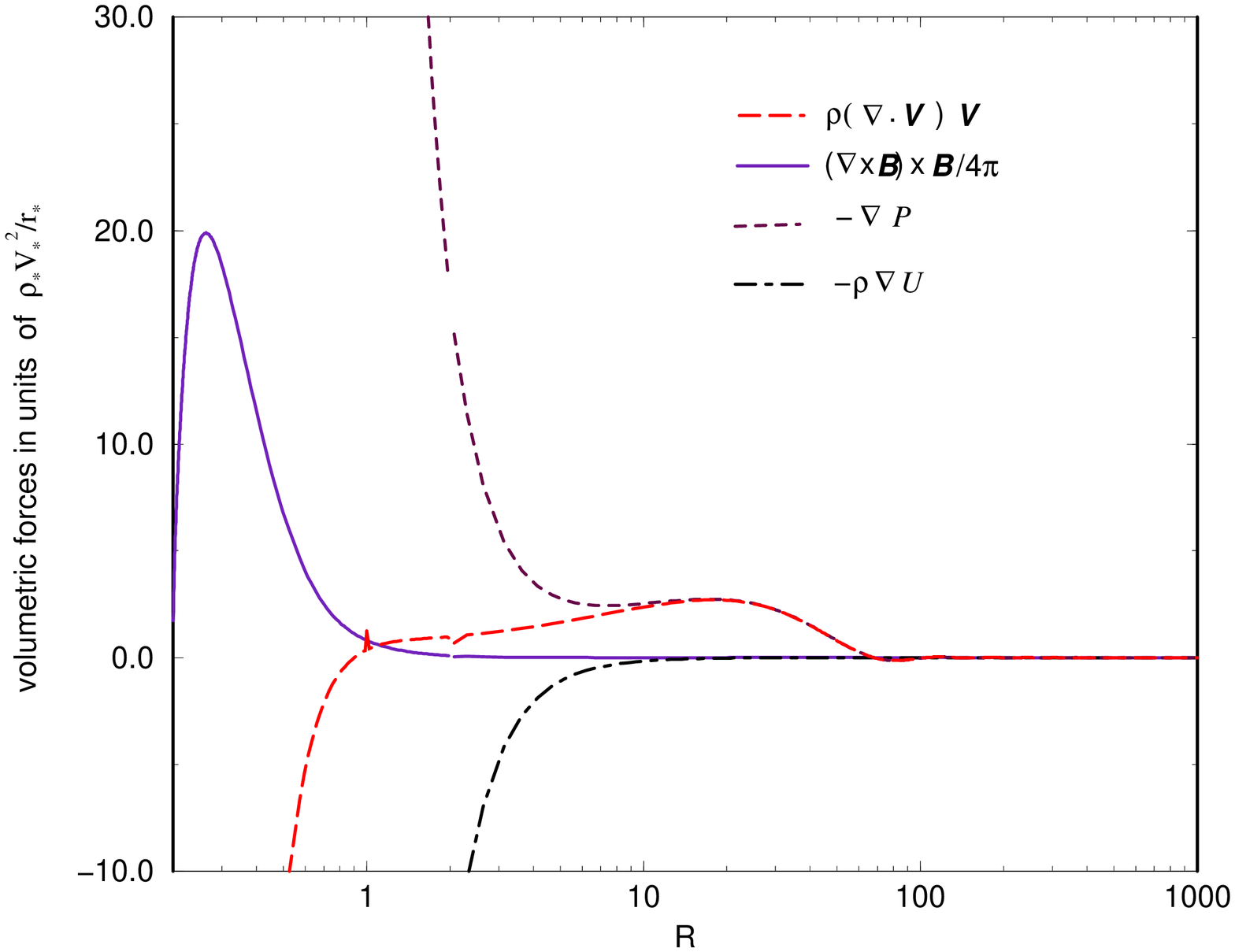,height=8.0truecm,width=8.0truecm,angle=270}}
%figure 5
\begin{figure*}
\centerline{\hspace{-0.5cm}\box1\hspace{-0.5cm}\box2}
\caption{\label{perp/para_a}
In the left panel are plotted the components of the magnetic (solid), pressure gradient 
(small dashes), gravitational (dot-dashed) and total acceleration (long dashes) 
perpendicular to the poloidal streamlines on line $\alpha=\alpha_{lim}$ 
for the parameters of the previous figure. In the the right panel the
corresponding components parallel to the poloidal lines are plotted  
also for {\bf case (a)} and the same parameters: 
$\epsilon=0.5$, $\xi=10$, $\delta \nu^2=3.5$, $\delta_0 \nu^2=0.1$, 
$\mu_0=0$, $F_{\star}=1$ and $p_{\star} \approx 2.2655 $}
\end{figure*}
%figure 6
\begin{figure*}
\centerline{
\psfig{file=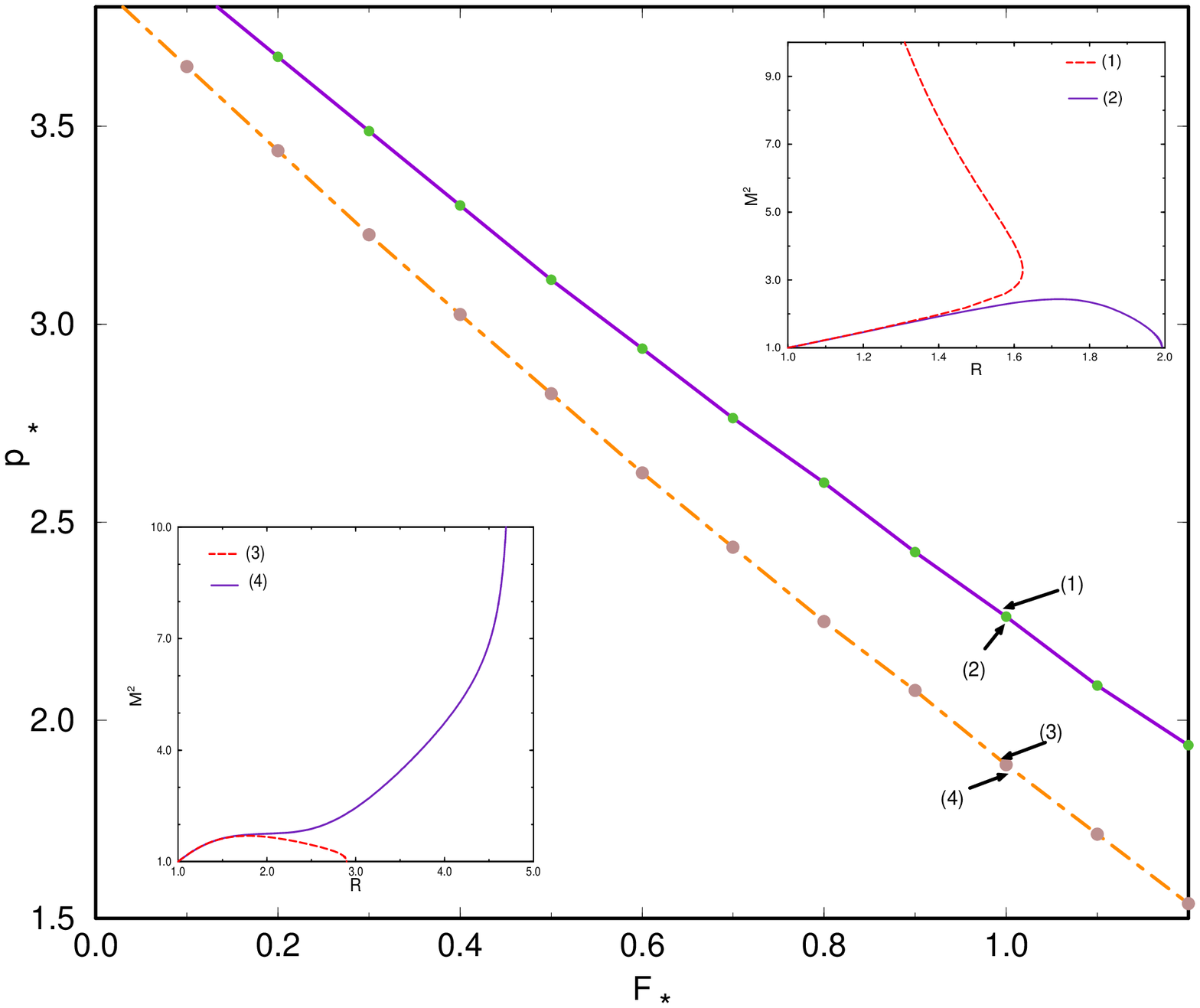,height=10.0truecm,width=19.5truecm,angle=180}}
\caption{\label{topo_a}
The solid line gives the relation between the expansion factor $F_{\star}$ 
and the slope $p_{\star}$ of $M^2(R)$ at the Alfv\'en point for a solution 
through all critical points, for {\bf case (a)} with parameters: 
$\epsilon=0.5$, $\xi=10$, $\delta \nu^2=3.5$, $\delta_0 \nu^2=0.1$, 
$\mu_0=0$. 
The topologies of $M^2(R)$ at the neighboring points (1), (2), (3) and (4) 
are also shown.} 
\end{figure*}
In all these cases we have $\xi (M^2_{\infty} -G_{\infty}^{4})>0$, or,  
$\xi \left( \left({\rho V_r^2}/{2}\right)_{\alpha=0,R\gg1}-\left(
{\rho V_r^2}/{2}\right)_{\alpha=0,R=1}\right)>0$.
In other words the sign of $\xi$ determines if the poloidal kinetic
energy on the axis is larger at the Alfv\'en point or at infinity.
Thus, according to the range of values of  $\xi$ and $\epsilon$ 
we distinguish the following cases:

\subsection{Case (a): $0<\epsilon<1$, $\xi>0$}

In this case cylindrical asymptotics is achieved through 
small amplitude 
oscillations of 
decaying amplitude (Type I solutions). In the left panel of 
Fig. (\ref{stream_a}) 
the shape of the field/streamlines on the poloidal plane is shown in the 
inner region between the stellar base, the Alfv\'en (dashed, $R=1$) and fast 
(dot-dashed, $R=2$) critical surfaces. The poloidal lines are almost radial 
up to the Alfv\'en surface while after the fast critical surface they 
have attained a cylindrical shape. However, the final cylindrical shape of 
the poloidal field/stream lines is reached further out, i.e., at about $R=20$, 
as it is shown in the larger scale of Fig. (\ref{stream_a}) 
(right panel) where  
their asymptotically cylindrical shape can be better seen. The bending of 
the poloidal field/stream lines towards the magnetic/rotational axis 
is caused by the magnetic pinching force as it can be seen in the left 
panel of Fig. (\ref{perp/para_a}) where the various components of 
the forces acting on the plasma perpendicular to the poloidal fieldlines 
are plotted. 
In the inner region of the outflow, the total inertial force 
perpendicular to the lines (centripetal force) is almost exclusively 
provided by the inwards magnetic force, with the outwards pressure gradient  
balancing the inward component of gravity. Asymptotically however, the 
magnetic pinching force and gravity are neglegible and 
the pressure gradient of the underpressured jet balances the 
centrifugal force. 
The acceleration of the plasma along the poloidal lines can be seen in 
the right panel of Fig. (\ref{perp/para_a}). 
Evidently, in the inner region gravity is balanced 
by the pressure gradient force and the plasma is accelerated only by the 
remaining magnetic force while in the outer region where gravity and the 
magnetic force are negligible, it is accelerated by the dominant pressure 
gradient force.  As it may also seen in Fig. (\ref{vel_a}) most of the 
acceleration occurs on the far region at $R \sgreat 10$ by the thermal pressure 
gradient force.

\setbox1=\vbox{\hsize=11truecm\vsize=8truecm
\psfig{figure=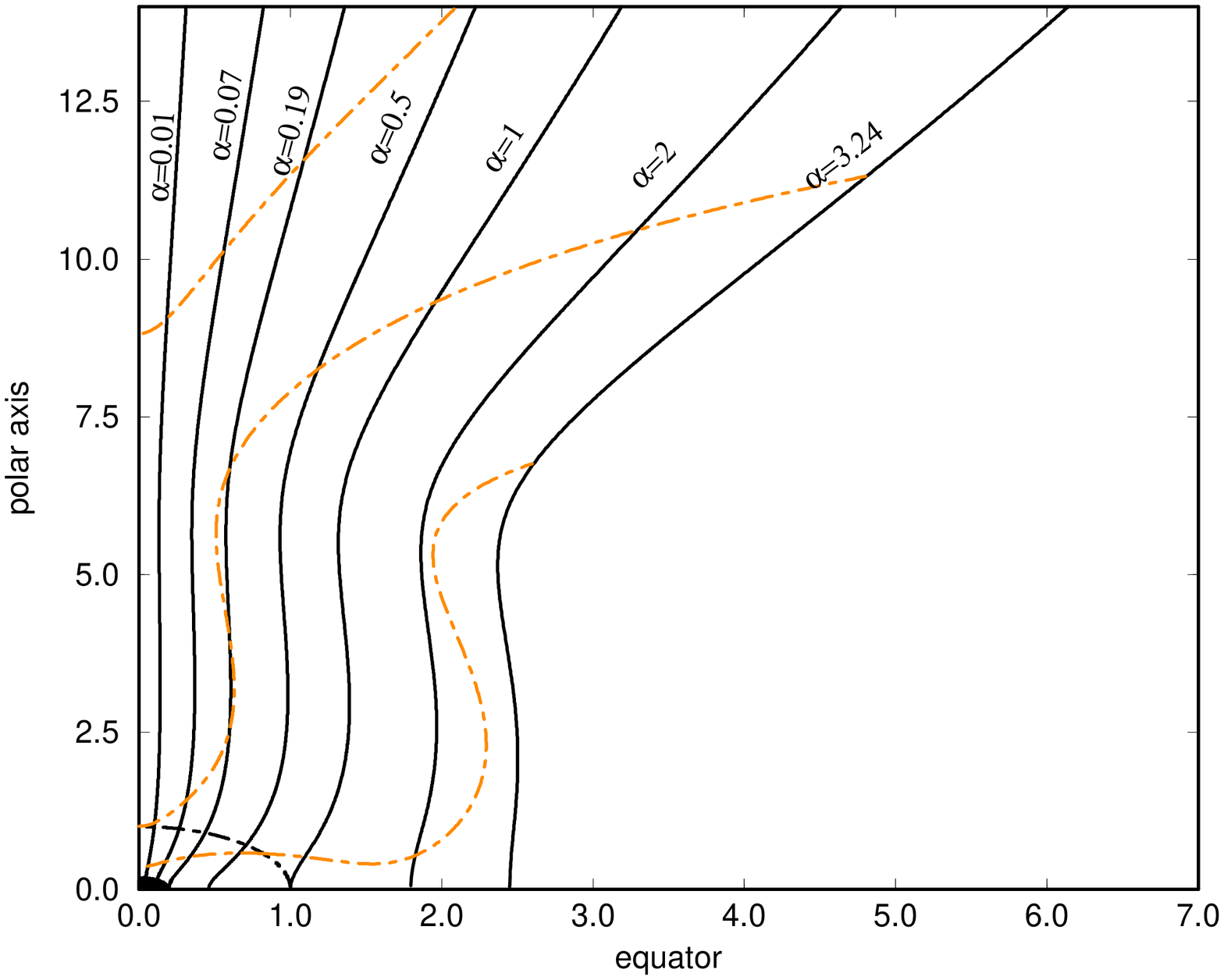,height=8.0truecm,width=8truecm,angle=270}}
\setbox2=\vbox{\hsize=11truecm\vsize=8truecm
\psfig{figure=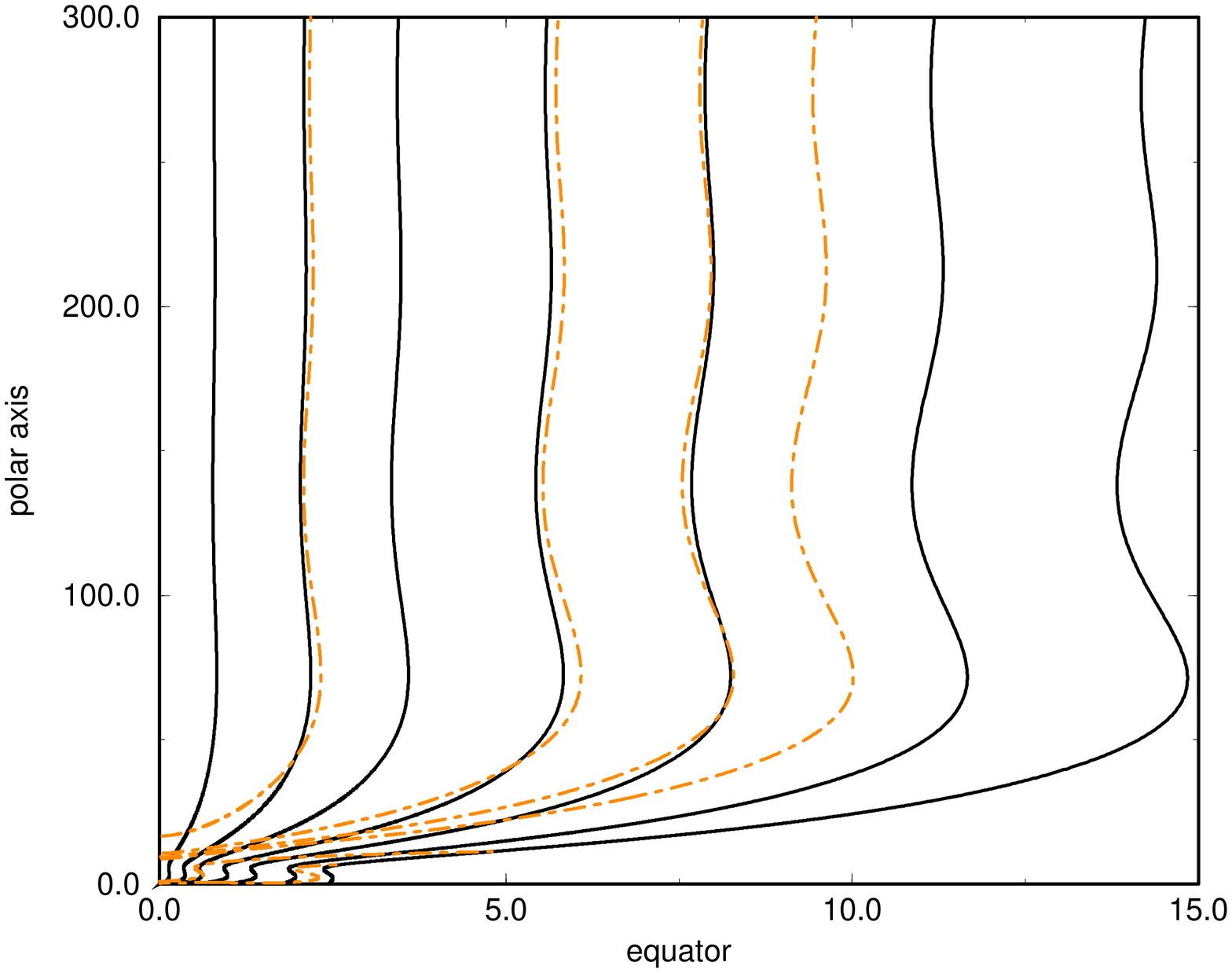,height=8truecm,width=8truecm,angle=270}}
%figure 7
\begin{figure*}
\centerline{
%\hspace{-.5cm}
\box1\hspace{-4.cm}\box2}
\caption{\label{stream_b}
Poloidal streamlines close and far from the central object
for {\bf case (b)} with parameters: 
$\epsilon=0.5$, $\xi=-5$, $\delta \nu^2=4$, $\delta_0 \nu^2=0.001$, 
$\mu_0=0$, $F_{\star}=1$ and $p_{\star} = 2$. With dotted lines the density 
isocontoures are indicated with $\rho/\rho_{\star}=0.1,1,10$  
from top to bottom in the left panel and $\rho/\rho_{\star}=0.01,0.04,0.07,1,10$ from 
left to right in the right panel.
The Alfv\'en surface in the left panel is indicated by dashed lines.}
\end{figure*}
%figure 8
\begin{figure}
\centerline{\psfig{file=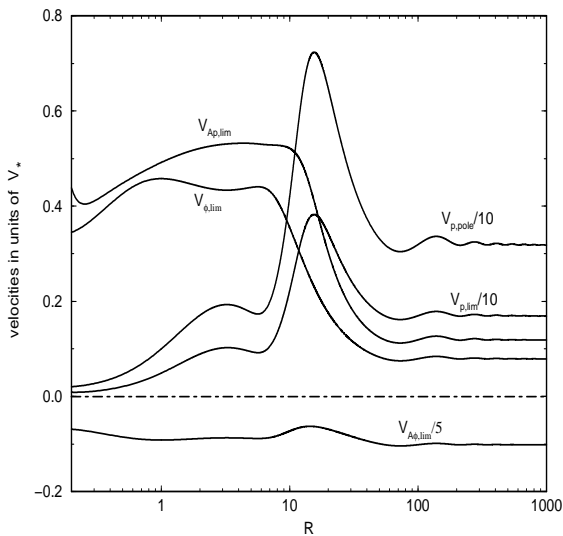,height=8.5truecm,width=8.5truecm,angle=270}}
\caption{\label{vel_b}
Dimensionless velocities for case ({\bf b}) with parameters: 
$\epsilon=0.5$, $\xi=-5$, $\delta \nu^2=4$, $\delta_0 \nu^2=0.001$, 
$\mu_0=0$, $F_{\star}=1$ and $p_{\star} = 2$}
\end{figure}
The solution discussed in this representative example crosses the modified by 
self-similarity fast critical point and a note is in order here on how 
such a solution may be obtained. First, we integrate Eqs. (\ref{dfdr}) -- 
(\ref{dp0dr}) downstream of the Alfv\'en critical point at which $R=G=M=1$,  
$F=F_{\star}$, $P_1=P_{1 \star}$ and $P_0=P_{0 \star}$, 
a free parameter which determines the 
pressure at infinity. At $R=1$ the Alfv\'en regularity condition 
relates $F_{\star}$, $p_{\star}$ and $P_{1 \star}$, 
Eq. ({\ref{alfcrit}}).
%Hence, for a given $P_{1 \star}$ and for each value of $F_{\star}$ 
%there is only one value of the Alfv\'en 
%number slope $p_{\star}$ 
Also there is a relation between $F_{\star}$,$p_{\star}$
such that the solution passes through the fast 
critical point; this is the solid line in Fig. (\ref{topo_a}).
Assume for example that we choose $F_{\star}=1$ and we vary $p_{\star}$, 
Fig. (\ref{topo_a}). 
There is only one value of $p_{\star} \approx 2.26$ which satisfies the 
Alfv\'en regularity condition and the solution crosses the fast critical 
point.  
For other values of $p_{\star}$ above and below $p_{\star} \approx 2.26$ 
we have three different types of unphysical solutions shown in  
Fig. (\ref{topo_a}):
\begin{itemize}
\item
from point (1) of Fig. (\ref{topo_a}) corresponding to $p_{\star}$ 
higher than $2.26$ we get solutions in which the 
denominator of the differential equation for $M^2$  becomes zero and the 
curve $M^2(R)$ turns back to smaller distances, 
\item
from point (2) of Fig. (\ref{topo_a}) corresponding to $p_{\star}$ 
lower than $2.26$ till point (3) we get solutions in which the 
numerator of the differential equation for $M^2$ becomes zero and then the 
solutions become again subAlfv\'enic,
\item
finally, from point (4) of Fig. (\ref{topo_a}) we get solutions in which 
there is a distance $R$ wherein $M \rightarrow \infty $ and the solutions 
terminate there. 
\end{itemize}
A fine tuning between points (1) and (2) gives the unique solution which goes 
to infinity with superAlfv\'enic and superfast radial velocity,  
satisfying also the causality principle for the propagation of MHD 
perturbations.
After finding such a critical value for $p_{\star}$ we also integrate 
Eqs. (\ref{dfdr}) - (\ref{dp0dr}) upstream of the Alfv\'en point.
%\vfill
%\vspace*{\fill}
%\newpage

\subsection{Case (b): $0<\epsilon<1$, $\xi<0$ }

\begin{figure*}
\centerline{
\psfig{file=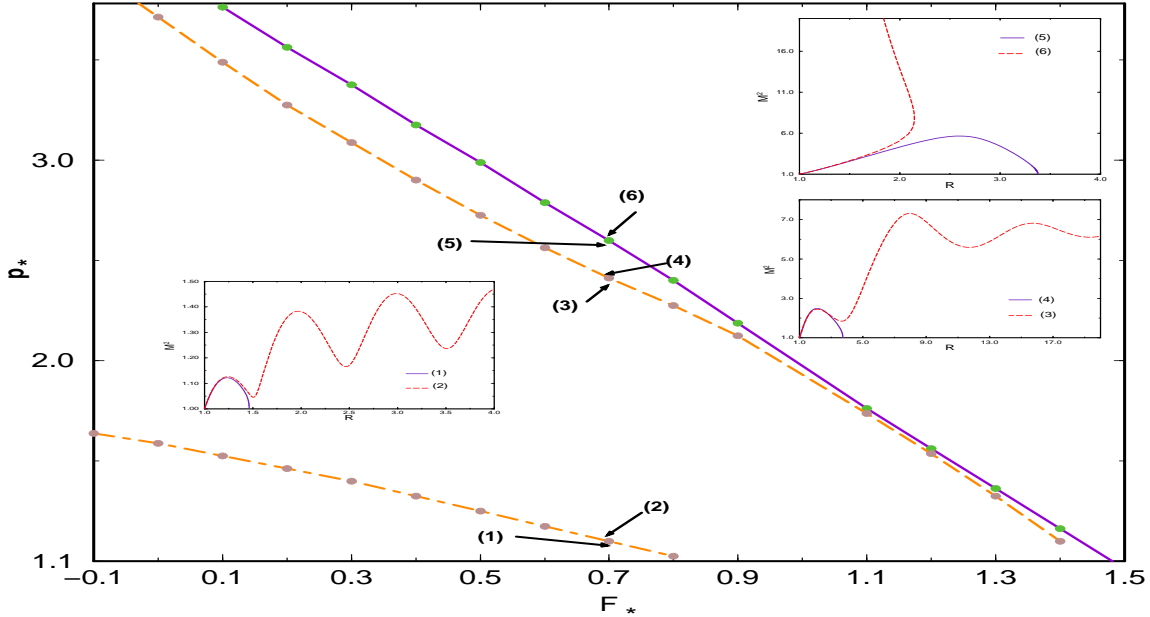,height=10.0truecm,width=18.5truecm,angle=180}}
\caption{\label{topo_b}
The solid line gives the relation between the expansion factor $F_{\star}$ 
and the slope $p_{\star}$ of $M^2(R)$ at the Alfv\'en point for a solution 
through all critical points, for {\bf case (b)} with parameters: 
$\epsilon=0.5$, $\xi=-5$, $\delta \nu^2=4$, $\delta_0 \nu^2=0.001$, 
$\mu_0=0$. 
The topologies of $M^2(R)$ at the neighboring points (1) to (6) 
are also shown.} 
\end{figure*}

In this case we may have two possibilities. In one the solution crosses the 
fast critical point and the situation is similar to the previous case (a). 
At the same time however asymptotically cylindrical solutions exist which 
do not cross the modified fast critical point, beeing simply superAlfvenic. 
An example of this type of behaviour is shown in Figures (\ref{topo_b}) - 
(\ref{vel_b}). 
As in case (a), cylindrical asymptotics is achieved through 
oscillations of decaying amplitude (Type I solutions). 
In the left panel of Fig. (\ref{stream_b})  
the shape of the field/streamlines on the poloidal plane is shown in the 
inner region between the stellar base and the Alfv\'en (dashed, $R=1$) 
critical surface. The poloidal lines are almost radial 
up to this Alfv\'en surface while outside $R=1$ they 
attain a cylindrical shape. However, the final cylindrical shape of 
the poloidal field/stream lines is reached further out, i.e., at about $R=20$, 
as it is shown in the larger scale of the right panel of 
Fig. (\ref{stream_b}) where  
their asymptotically cylindrical shape obtained through the decaying amplitude 
oscillations can be better seen. 
%figure 9
As in case (a), the focusing of the poloidal field/stream lines towards 
the magnetic and rotation axis is caused predominantly by the magnetic 
pinching force; this may be seen in the left panel of Fig. (\ref{perp/para_b}) 
where the various components of 
the forces acting on the plasma {\it{perpendicular}} to the poloidal fieldlines 
are plotted. In the inner region of the outflow $R$\sles $1$, the total inertial 
force perpendicular to the lines (centripetal force) is almost exclusively 
provided by the inwards magnetic force. 
In the far zone where gravity is negligible, $R$ \sgreat $1$, the inwards magnetic 
pinching force is balanced by the pressure gradient of the 
overpressured jet and the centrifugal force.   
The acceleration of the plasma {\it{along}} the poloidal lines can be seen in 
the right panel of 
Fig. (\ref{perp/para_b}). In the inner region $R$ \sles $1$ the magnetic and pressure gradient 
forces accelerate the plasma; in the outer region where gravity and the 
magnetic forces are negligible, the pressure gradient force is left alone 
to accelerate the plasma. 
As in case (a), it may also be seen in the right panel of Fig. (\ref{perp/para_b}) 
that most of the acceleration occurs in the far region at $R\approx 10$ by 
the thermal pressure gradient force. \\
Figure (9) is a plot of the values of $p_{\star}$ and $F_{\star}$ 
for which the fast point is crossed.
As in case (a), we integrate Eqs. (\ref{dfdr}) - 
(\ref{dp0dr}) downstream of the Alfv\'en critical point at which $R=G=M=1$,  
$F=F_{\star}$, $P_1=P_{1 \star}$ and $P_0=P_{0 \star}$. At $R=1$ the Alfv\'en regularity condition 
relates $F_{\star}$, $p_{\star}$ and $P_{1 \star}$, 
Eq. (\ref{alfcrit}).
%Hence, for a given $P_{1 \star}$ and for each value of $F_{\star}$ there is only one value of the Alfv\'en 
%number slope $p_{\star}$ 
Also there is a relation between $F_{\star}$,$p_{\star}$
such that the solution passes through the fast 
critical point; this is the solid line in Fig. (\ref{topo_b}).
Assume for example that we choose $F_{\star}=0.7$ and we vary $p_{\star}$, 
Fig. (\ref{topo_b}). 
There is only one value of $p_{\star} \approx 2.6$ which satisfies the 
Alfv\'en regularity condition and the solution crosses the fast critical 
point.  

For other values of $p_{\star}$ above and below $p_{\star} \approx 2.6$ 
we have different types of solutions shown in  
Fig. (\ref{topo_b}):
\begin{itemize}
\item
at point (6) of Fig. (\ref{topo_b}) corresponding to $p_{\star}$ 
higher than $p_{\star} \approx 2.6$ we get solutions in which the 
denominator of the differential equation for $M^2$  becomes zero and the 
curve $M^2(R)$ turns back to smaller distances, 
\item
at points (5), (4) and (1) of Fig. (\ref{topo_b}) we get solutions in which the 
numerator of the differential equation for $M^2$ becomes zero and then the 
solutions become again subAlfv\'enic,
\item
at points (2) and (3) of Fig. (\ref{topo_b}) we get oscillatory solutions 
wich do not cross the fast critical point.
These solutions were shown in figures (\ref{stream_b}).  
\end{itemize}
A fine tuning between points (6) and (5) gives the unique solution which goes 
to infinity with superAlfv\'enic and superfast radial velocity. \\

Note that in this case there exists a value $\alpha_{out}$ where 
$V_{\phi}=0$ and $B_{\phi}=0$. In this streamline the enclosed poloidal current is 
zero. For $\mu=9$ and the parameters as in Fig. (\ref{stream_b}), 
$\alpha_{out}=3.24$.
\\
\setbox1=\vbox{\hsize=8.0 truecm \vsize=8.0truecm
\psfig{figure=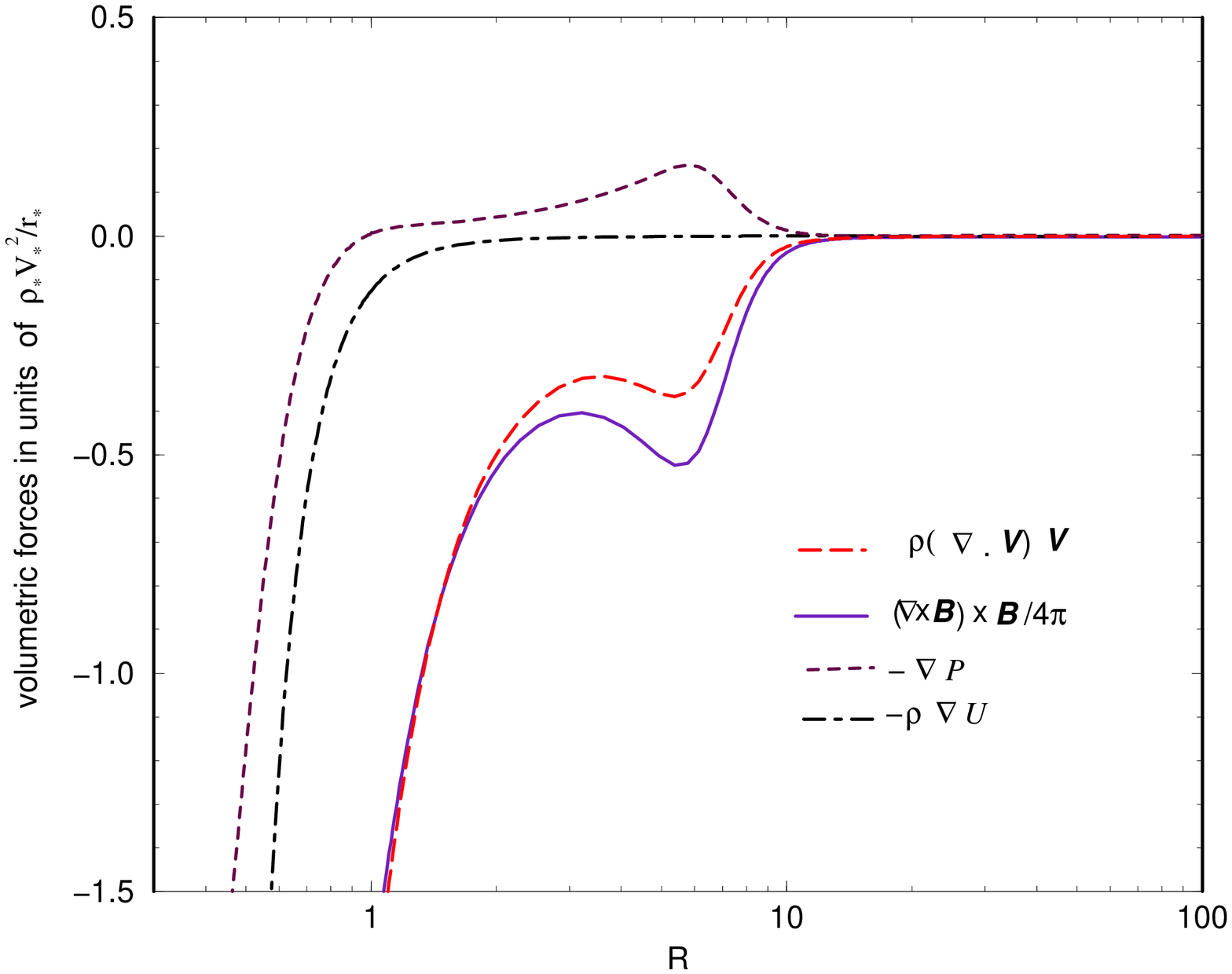,height=8.0truecm,width=8.0truecm,angle=270}}
\setbox2=\vbox{\hsize=8.0 truecm \vsize=8.0truecm
\psfig{figure=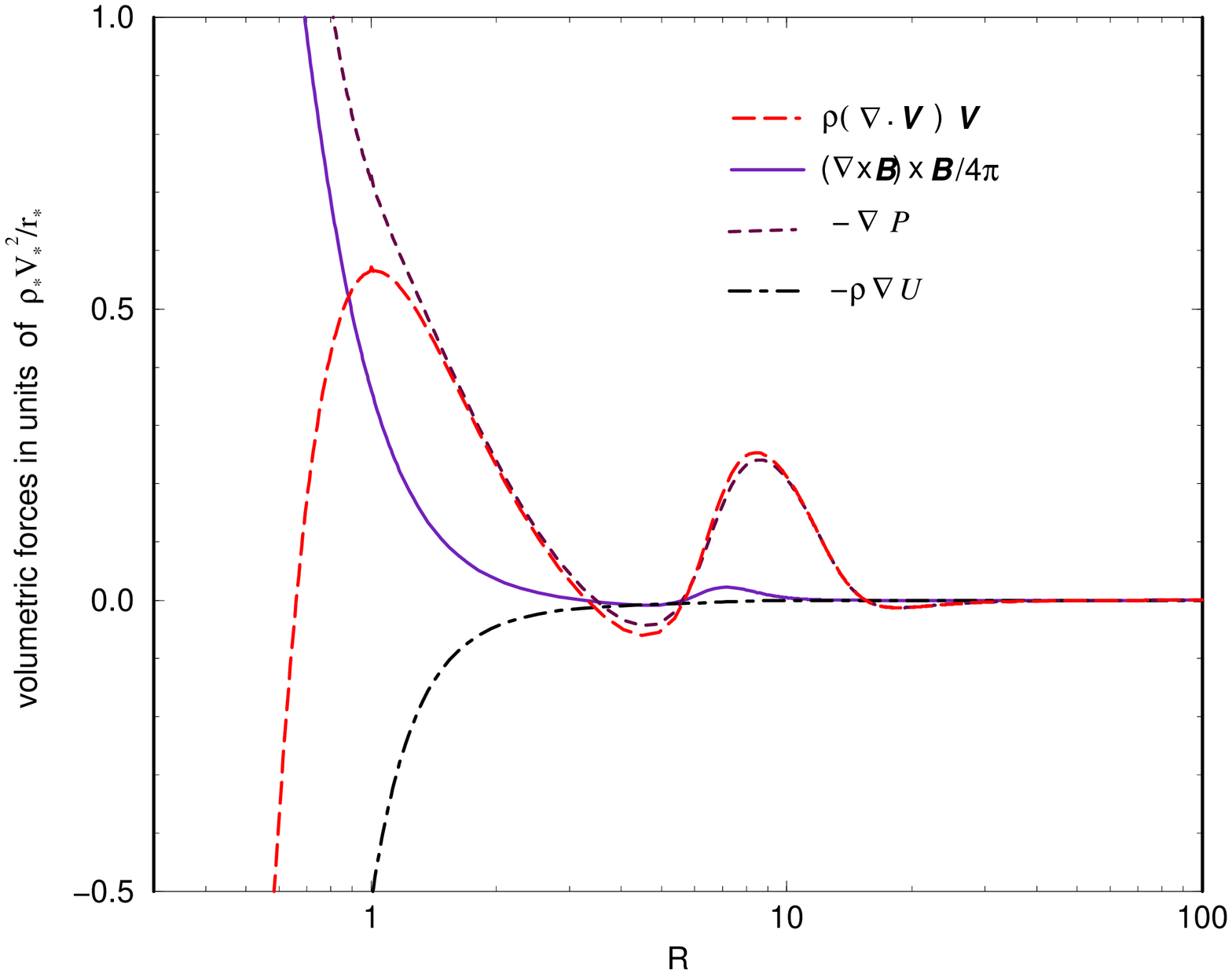,height=8.0truecm,width=8.0truecm,angle=270}}
%figure 10
\begin{figure*}
\centerline{\hspace{-0.5cm}\box1\hspace{-0.5cm}\box2}
\caption{\label{perp/para_b}
In the left panel are plotted the components of the magnetic (solid), pressure gradient 
(small dashes), gravitational (dot-dashed) and total acceleration (long dashes) 
perpendicular to the poloidal streamlines on line $\alpha=\alpha_{lim}$. 
In the right panel the corresponding components parallel to the poloidal lines are plotted  
also for {\bf case (b)} and the same set of parameters: 
$\epsilon=0.5$, $\xi=-5$, $\delta \nu^2=4$, $\delta_0 \nu^2=0.001$, 
$\mu_0=0$, $F_{\star}=1$ and $p_{\star}= 2 $}
\end{figure*}
\setbox1=\vbox{\hsize=11truecm\vsize=8truecm
\psfig{figure=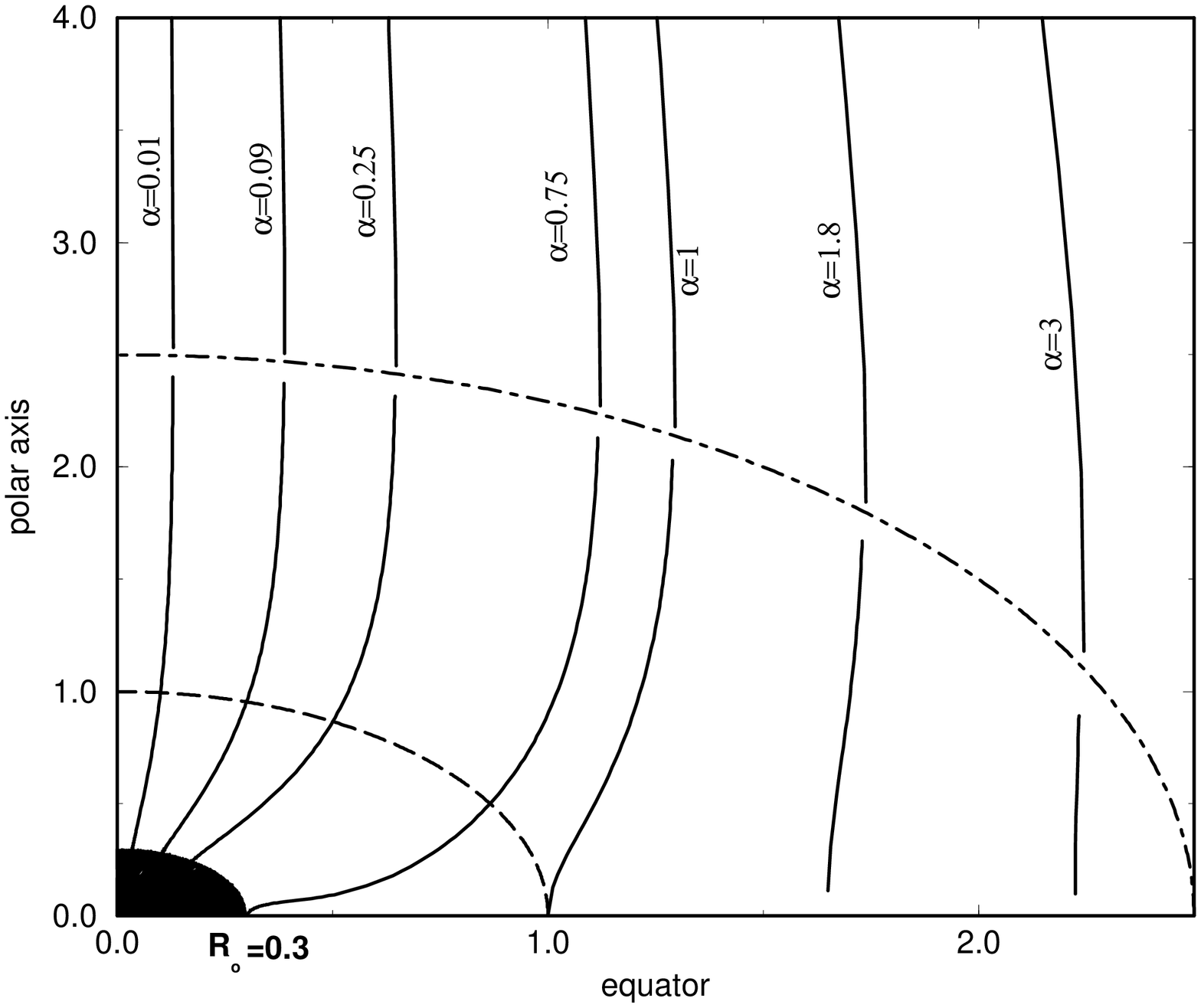,height=8truecm,width=8truecm,angle=270}}
\setbox2=\vbox{\hsize=11truecm\vsize=8truecm
\psfig{figure=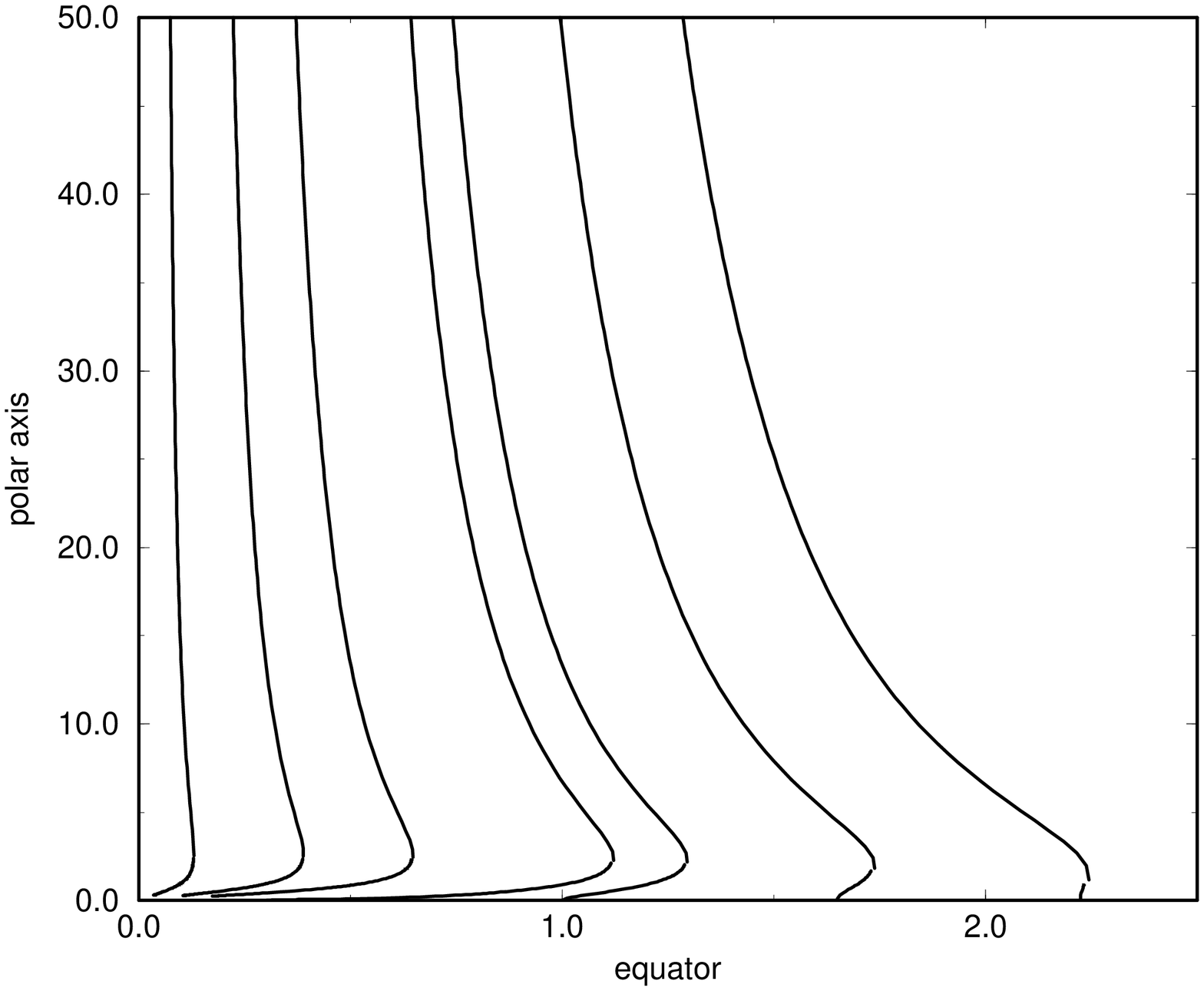,height=8truecm,width=8truecm,angle=270}}
%figure 11
\begin{figure*}
\centerline{
%\hspace{-1.cm}
\box1\hspace{-4.cm}\box2}
\caption{\label{stream_c}
Poloidal streamlines close and far from the central object
for {\bf case (c)} with parameters: 
$\epsilon=2$, $\xi=10$, $\delta \nu^2=4$, $\delta_0 \nu^2=0.1$, 
$\mu_0=0$, $F_{\star}=0.8$ and $p_{\star} \approx 2.5636$.
The Alfv\'en (fast) surface is indicated by dashed (dot-dashed) lines.}
\end{figure*}

\subsection{Case (c): $\epsilon>1$,  $\xi>0$ }
%figure 12
\begin{figure*}
\centerline{
\psfig{file=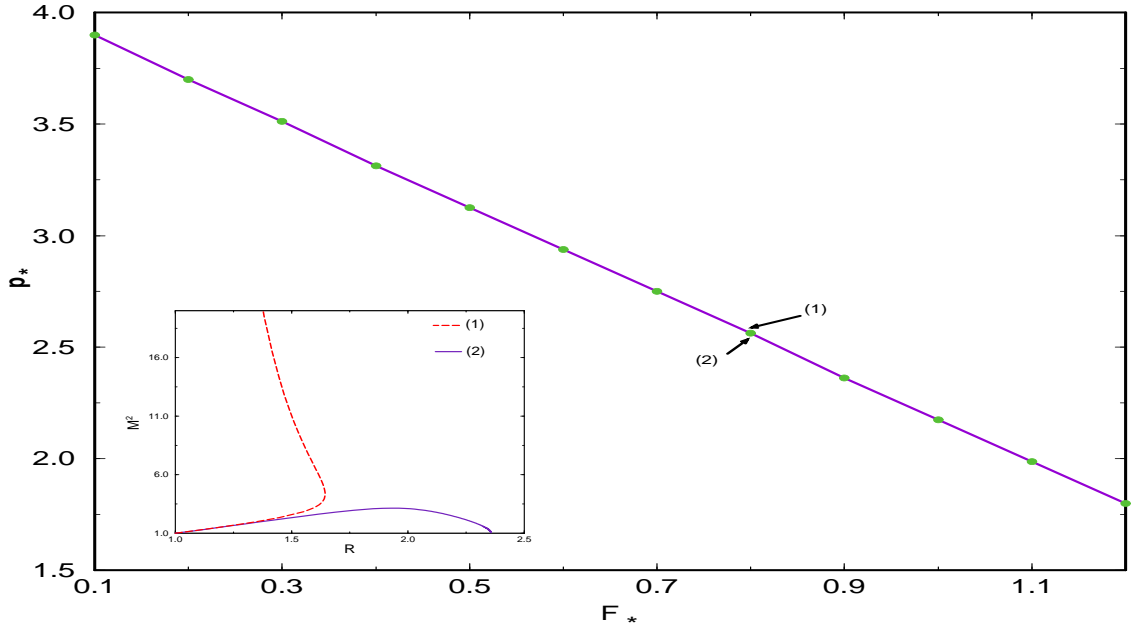,height=10.0truecm,width=18.5truecm,angle=180}}
\caption{\label{topo_c}
The solid line gives the relation between the expansion factor 
$F_{\star}$ and the slope $p_{\star}$ of $M^2(R)$ at the Alfv\'en point for 
a solution through all critical points, for {\bf case (c)} with parameters: 
$\epsilon=2$, $\xi=10$, $\delta \nu^2=4$, $\delta_0 \nu^2=0.1$, 
$\mu_0=0$.}
\end{figure*}

%figure 13
\begin{figure}
\centerline{\psfig{file=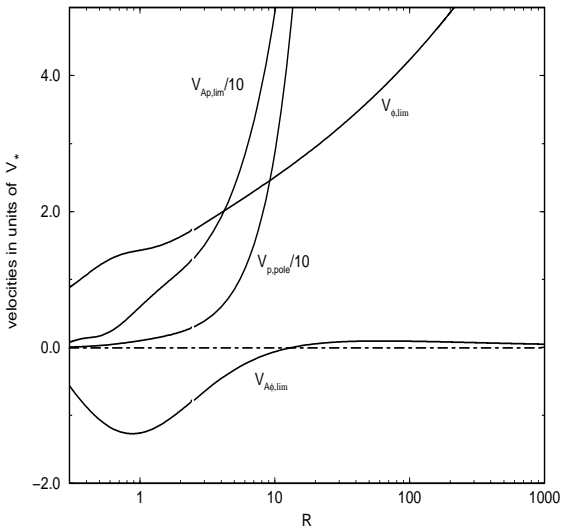,height=8.5truecm,width=8.5truecm,angle=270}}
\caption{\label{vel_c}
Dimensionless velocities for case ({\bf c}) with parameters: 
$\epsilon=2$, $\xi=10$, $\delta \nu^2=4$, $\delta_0 \nu^2=0.1$, 
$\mu_0=0$, $F_{\star}=0.8$ and $p_{\star} \approx 2.5636$.}
\end{figure}

As discussed in the beginning of Sec. 5, when $\epsilon >1$ the strong 
magnetic pinching force results in a jet of zero asymptotic radius; 
in addition, this asymptotics is achieved without oscillations, 
i.e., we obtain type II solutions, Fig. (\ref{topo_c}), (\ref{stream_c}) and 
(\ref{vel_c}). 
The values of $p_{\star}$ and $F_{\star}$ for which the solution crosses 
the fast critical point are shown in Fig. (\ref{topo_c}).

As with the previous cases, for each value of $F_{\star}$ there is only 
one value of the Alfv\'en 
number slope $p_{\star}$ such that the solution passes through the fast 
critical point; this is the solid line in Fig. (\ref{topo_c}).
Assume for example that we choose $F_{\star}=0.8$ and we vary $p_{\star}$, 
Fig. (\ref{topo_c}). 
There is only one value of $p_{\star} \approx 2.53$ which satisfies the 
Alfv\'en regularity condition and the solution crosses the fast critical 
point.  
For other values of $p_{\star}$ above and below $p_{\star} \approx 2.53$ 
we have two different types of unphysical solutions shown in  
Fig. (\ref{topo_c}):

\begin{itemize}
\item
at point (1) of Fig. (\ref{topo_c}) corresponding to $p_{\star}$ 
higher than $2.53$ we get solutions in which the 
denominator of the differential equation for $M^2$  becomes zero and the 
curve $M^2(R)$ turns back to smaller distances, 
\item
at point (2) of Fig. (\ref{topo_c}) corresponding to $p_{\star}$ 
lower than $2.53$ we get solutions in which the 
numerator of the differential equation for $M^2$ becomes zero and then the 
solutions become again subAlfv\'enic,
\end{itemize}

A fine tuning between points (1) and (2) gives the unique solution which goes 
to infinity with superAlfv\'enic and superfast radial velocity.  
%After finding such a critical value for $p_{\star}$ we also integrate 
%Eqs. (\ref{dfdr}) - (\ref{dp0dr}) upstream of the Alfv\'en point.
Nevertheless, the jet radius goes to zero in this case.

%\vspace*{\fill}
%\newpage
\begin{table*}
\centering
%\begin{minipage}{140mm}
\caption{Astrophysical applications for cases (a) and (b)}
\begin{tabular}{@{}lcrrrrcrrr@{}}
%\begin{tabular}{l|rrrr|rrr}
&&\multicolumn{4}{c|}{case (a)}&
&\multicolumn{3}{c|}{case (b)}\\
&
& base
& Alfv\'en
& fast
& infinity
&
& base
& Alfv\'en
& infinity \\
\hline
$r(cm)$
&
& $2 \times 10^{11}$
& $ 10^{12} $
& $ 2.04 \times 10^{12} $
& $\gg r_{\star}$
&
& $2 \times 10^{11}$
& $ 10^{12} $
&$\gg r_{\star}$\\
$V_r(cm/sec)$
&
& $3\times 10^{4}$
& $ 7.7 \times 10^5$
& $ 1.8 \times 10^6$
& $5\times 10^{7}$
&
& $6.5 \times 10^6$
& $3.2 \times 10^7$
&$10^{8}$\\
$B_r(G)$
&
& $7\times 10^{-3}$
& $5.7 \times 10^{-4}$
& $4 \times 10^{-4}$
& $10^{-3}$
&
& $0.3$
& $6.2 \times 10^{-2}$
& $10^{-3}$\\
$\rho(gr/cm^3)$&
&
$1.6 \times 10^{-17}$
& $5.1 \times 10^{-2}$
& $1.5 \times 10^{-2}$
& $1.66 \times 10^{-21}$
&
& $8 \times 10^{-18}$
& $3.4 \times 10^{-19}$
&$1.66 \times 10^{-21}$\\
$M^2$
& $\,$
& $3.2 \times 10^{-3}$
& $1$
& $3.34$
& $31.3$
& $\,$
& $4.22 \times 10^{-2}$
& $1$
& $203$
\\
$G^2$
& $\,$
& $8.2 \times 10^{-2}$
& $1$
& $1.41$
& $0.477$
& $\,\,$
& $0.206$
& $1$
& $64$
\\
\hline
\end{tabular}
%\end{minipage}
\end{table*}

\section{Astrophysical Applications}

It should be noted that the purpose of this paper has not been to construct 
a specific model for a given collimated outflow; 
instead, our goal has been to outline via a specific class of exact and 
self-consistent models, the 
interplay of the various MHD processes contributing into the acceleration 
and collimation of jets. 
Nevertheless, the illustrative examples analysed in this paper 
can be compared with the observable characteristics of outflows   
from stellar or galactic objects, say, those associated with young 
stellar objects. For this 
purpose, in the following we establish the connection between the 
nondimensional models and the observable parameters of the outflow. \\

Suppose that at the polar direction of the stellar surface 
($r=r_0\,,\alpha=0$) 
we know the values of $V_r \,, B_r$ and $\rho$, say, $V_0$, $B_0$ and 
$\rho_0$, respectively, such that we calculate 
%$M_0=\frac{V_0 \sqrt{4 \pi \rho_0}}{B_0}$.  
$M_0=V_0 \sqrt{4 \pi \rho_0}/B_0$.  
From the integration we can find the distance $R_0$ where $M(R_0)=M_0$.
Thus, we may calculate the Alfv\'en distance $r_{\star}=r_0/R_0$.
Each line which has its footpoint on the stellar surface at angle $\theta_i$ 
is labeled by $\alpha=\left({r_0 \sin \theta_i}/{r_{\star} G\left(r_0/
r_{\star}\right)}\right)^2$. 
The last line originating from the star is $\alpha_{lim}$.
Each line which has its footpoint on the 
disk at distance $r_i>r_0$ from the axis of 
rotation is labeled by 
$\alpha=\left({r_i}/{r_{\star} G\left(
r_i/r_{\star}\right)}\right)^2> \alpha_{lim}$.\\
If at the stellar surface $G(R_0)=G_0$ we find the Alfv\'en values, 
$V_{\star}=V_0 G_0^2/M_0^2$, $B_{\star}=B_0 G_0^2$, $\rho_{\star}=
\rho_0 M_0^2$ and from Eqs. (\ref{density}) to (\ref{B}) we can find all 
physical quantities at any point.
For example at $R \gg 1$, $\alpha=0$ we have the following asymptotic values: 
$V_{\infty}=V_0 G_0^2 M_{\infty}^2/M_0^2 G_{\infty}^2$, 
$B_{\infty}=B_0 G_0^2 /G_{\infty}^2$, $\rho_{\infty}=\rho_0 
M_0^2/M_{\infty}^2\,.$\\

\subsection{Model of case (a)}

For a typical solution with parameters as those plotted in  Fig. (\ref{stream_a}),
the values of characteristic physical quantities are shown in Table 1.
These values refer to the intersection of the rotational axis with
(i) the stellar surface, (ii) the Alfv\'en singular surface,
(iii) the modified by the self-similarity fast singular surface and
(iv) infinite distance from the source.
For a solar type stellar mass $2\times 10^{33} gr$ we have $\nu^2=462$ 
while for $\mu=0.01$ the angular velocity at the equatorial point of the 
stellar surface has the solar value $2\times 10^{-6}/sec$ . \\
Note that in this case (a) the toroidal component of the magnetic field 
changes sign at some spherical surface 
({\it cf.} the velocity $V_{A \phi \,, lim}$ in 
Fig. (\ref{vel_a}). This means that the poloidal current enclosed by this surface 
is zero. All fieldlines which pass through this surface have the same 
cylindrical distance 
from the axis with the Alfv\'en point ($G=1$ at this spherical surface)
while for larger distances $G<1$.
After crossing this surface the Poynting flux 
changes its sign and thus the toroidal component of the velocity becomes 
large enough (because $V_{\phi}/\varpi \Omega=
\left(M^2-G^2\right)/G^2 \left(M^2-1\right)\approx 1/G^2$).
It is worth to note that even with such rather weak strengths of the
magnetic field, collimation is readily achieved.

\subsection{Model of case (b)}

For a typical solution with parameters as those plotted in  Fig. 
(\ref{stream_b}),
the values of characteristic physical quantities are also shown in Table 1 
and these values refer, as before, to the intersection of the rotational 
axis with (i) the stellar surface, (ii) the Alfv\'en singular surface and
(iv) infinite distance from the source.
The last line connected with the star has $\alpha_{lim}=0.19$ while the 
disk has a radius $2.45 \times 10^{12}cm$.
If we choose a one solar mass star, $\nu^2=0.3$ while for $\mu=9$,
the stellar equator rotates with a speed $2.7 \times 10^4 cm/sec$ (the 
angular velocity is $1.3 \times 10^{-7}/sec$).\\

The asymptotic radius of the jet (which is bounded with the line $\alpha_{out}$)
is $1$ $A.U.$ while the part of the flow starting from the stellar surface has 
a radius $0.23$ $A.U.$.This part of the jet is collimated at 
a distance of about $4$ $A.U.$ from 
the equatorial plane, while the whole solution collimates at a height 
of $3.3$ $A.U.$ 
These results are consistent with recent observations of YSO's (Ray et al 1996).
%figure 14
\begin{figure}
\centerline{\psfig{file=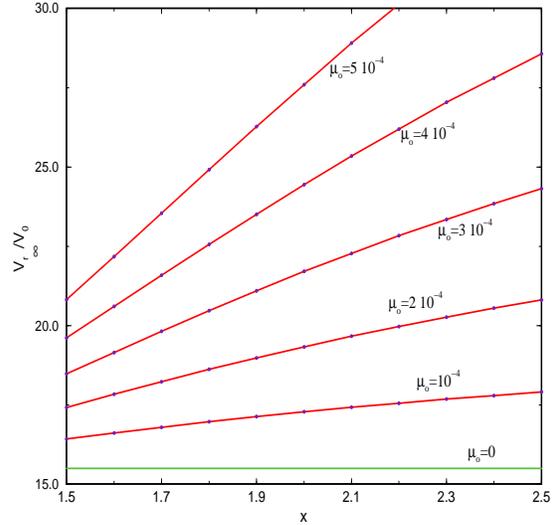,height=8.5truecm,width=8.5truecm,angle=270}}
\caption{\label{rad1}
Dimensionless asumptotic values of the radial velocity as a function 
of the radiative force parameters $x$ and $\mu_0$ for case ({\bf b}) 
with parameters: 
$\epsilon=0.5$, $\xi=-5$, $\delta \nu^2=4$, $\delta_0 \nu^2=0.001$, 
$\nu^2=0.3$, $F_{\star}=1$ and $p_{\star} = 2$. 
In all these cases in the surface of the star $M_0^2=4.22\times 10^{-2}$.}
\end{figure}
%figure 15
\begin{figure}
\centerline{\psfig{file=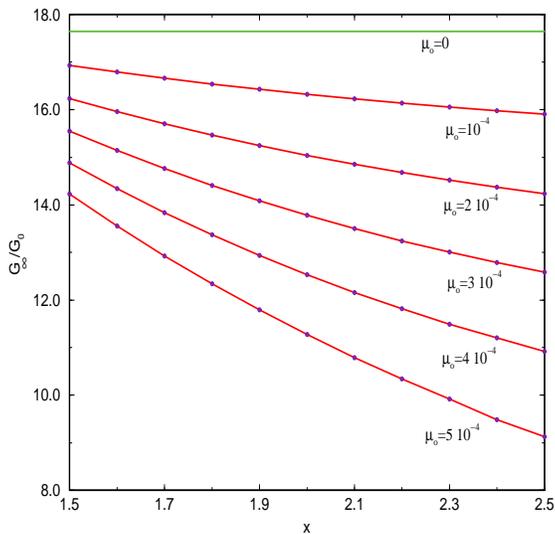,height=8.5truecm,width=8.5truecm,angle=270}}
\caption{\label{rad2}
Dimensionless asumptotic values of the radius of the jet 
as a function of the radiative force parameters $x$ and $\mu_0$ for case 
({\bf b}). The other parameters are the same as in the previous figure.}
\end{figure}

\subsection{Model of case (b) including radiation}

There are two parameters ($\mu_0 \,, x$) determining the radiative force 
(the third is included into $\nu^2$). 
For the parameters of case (b) but for $\mu_0 \neq 0$ we examine the 
effect of the radiative force on the velocity and the assymptotic radius of 
the outflow.
As we expect, as the radiative force increases, the terminal velocity 
becomes larger, Fig. (\ref{rad1}) 
while the Alfv\'en surface moves closer to the stellar base. From mass 
conservation we expect that the cross sectional area of the jet decreases 
as $x$ increases, as it is shown in Fig. (\ref{rad1}).
%\vspace*{\fill}

\section{Summary and Conclusions}

In this paper we have examined a class of exact solutions of the set of 
the MHD equations (\ref{mhdforce}) - (\ref{mhdfluxes}) governing the 
kinematics of a magnetized outflow from a rotating gravitating object. 
For this system to be closed, an additional equation is needed to describe   
the energetics of the outflow, i.e., some form of the energy conservation 
principle, Eq. (\ref{firstlaw}). 
The often used simplifying polytropic relationship between 
pressure and density which corresponds to a specific functional form of the 
net heating/cooling in the plasma, was not used. This has the incovenient 
consequence that the sound speed is ill defined and it can be calculated only 
at the critical points. Besides this 
inconveniency we do not suffer any loss of generality in adopting a more 
general functional form of the total heating, than the polytropic assumption 
implies. As it was explained in Sec. 2, 
in both, the familiar polytropic case of constant $\gamma$ and the present 
nonconstant $\gamma$ approach, the detailed spatial distribution of the 
required heating can be calculated only a {\it posteriori}.  

The class of solutions which is analysed in this paper belongs to a group of 
nine classes of meridionally selfsimilar MHD solutions which have been shown to 
exist in the recent paper VT98 under the assumptions that the Alfv\'en Mach 
number is a function of the radial distance and the poloidal magnetic field 
has a dipolar angular dependence, Eq. (\ref{assumptions2}). These assumptions 
may be reasonable for outflows around the magnetic and symmetry axis of the 
system. No assumption was made about the asymptotics of the outflows. 
It is interesting that the self-consistently deduced shape of 
the streamlines and magnetic field lines was found to be helices wrapped 
on surfaces which asymptotically are cylindrical. In other words, the 
streamlines extended to infinite heights above the central object and its 
disk obtaining the form of a jet. It was shown that such collimation is 
obtained even with very weak magnitudes of the magnetic field.  
This result may be contrasted to the quite often refered Blandford  
\& Payne (1982) solutions which by overfocusing towards the axis terminate 
at finite heights above the disk. 
The new element the present model introduces in the self consistent modelling 
of MHD outflows is that it produces for the first time jet-type solutions 
extending from the stellar base to infinity and where the outflow 
crosses at a finite distance the fast critical point such that the 
MHD causality principle is satisfied.  
The cylindrical asymptotics of the present nonpolytropic solutions 
are consistent with the polytropic analysis of Heyvaerts \& Norman (1989) and 
also with the class of superAlfvenic but subfast at infinity solutions of Sauty 
\& Tsinganos (1994) for efficient magnetic rotators. 
However, no radial asymptotics was found in the present class of models,  
contrary to the other class of meridionally selfsimilar solutions 
examined in Sauty \& Tsinganos (1994) where for inefficient magnetic rotators
radial asymptotics has been found; it may be that the present model belongs  
to the group of efficient magnetic rotators.  

The topologies of the solutions are rather rich as it was shown in the 
plane defined by the slope of the Alfv\'en number $p_{\star}$ and the 
streamline expansion factor $F_{\star}$ at the Alfv\'en transition. 
For example, for a given streamline expansion factor $F_{\star}$  we 
obtained terminated solutions for $p> p_{\star}$, similarly to the 
corresponding terminated solutions in Parker's (1958) HD wind, or, the 
Weber \& Davis magnetized wind. For a given pressure at the Alfv\'en 
point, the requirement that a solution crosses the Alfv\'en and fast 
critical points eliminates the freedom in choosing $p_{\star}$
and $F_{\star}$ through the corresponding regularity and criticality 
conditions.  

A plotting of the various forces acting along and perpendicular to the 
poloidal streamlines reveals that the wrapping of the field lines 
around the symmetry axis is caused predominantly by the hoop stress of 
the magnetic field and it is already strong at the Alfv\'en (and fast)  
critical surface. Asymptotically the cylindrical column is confined by 
the interplay of the inwards magnetic pinching force, the outwards 
centrifugal force and the pressure gradient, as in Trussoni et al (1997). 
On the other hand, the acceleration of the plasma along the poloidal 
magnetic lines, in the near zone close to the Alfv\'en distance it is due 
to the combination of thermal pressure and magnetic forces while at the 
intermediate zone beyond the Alfv\'en point it is basically the pressure 
gradient that is responsible for the acceleration.   

\section*{Acknowledgments}

This research has been supported in part by a grant from the General 
Secretariat of Research and Technology of Greece. We thank J. Contopoulos, 
C. Sauty, G. Surlantzis and E. Trussoni for helpful discussions. 

%\vspace*{\fill}

%\newpage
%BIBLIOGPAPHY

%\newpage
\appendix
\section[]{functions of $R$} 

\begin{equation}\label{F}
F=2-R\frac{G^{2'}}{G^2}
\equiv  \frac{\partial \ln \alpha (R, \theta) }{\partial \ln R}
\,.
\end{equation}
\begin{equation}\label{f1}
f_1=-\frac{1}{G^4}
\,,
\end{equation}
\begin{equation}\label{f2}
f_2=-\frac{F^2-4}{4G^2 R^2}
\,,
\end{equation}
\begin{equation}\label{f3}
f_3=-\frac{1}{G^2}\left(\frac{1-G^2}{1-M^2}\right)^2
\,,
\end{equation}
\begin{equation}\label{f4}
f_4=\frac{F }{2RG^2}M^{2'}-\frac{1-M^2}{2RG^2}F^{'}-\frac{\left(1-M^2\right)F\left(F-2\right)}{4R^2 G^2}
\,,
\end{equation}
with,
\begin{equation}\label{F'}
F'=\frac{F}{1-M^2}M^{2'}-\frac{F\left(F-2\right)}{2R}-\frac{2RG^2}{1-M^2}f_4
\,,
\end{equation}
\begin{equation}\label{f5}
f_5=\frac{G^4-M^2}{G^2M^2\left(1-M^2\right)}
\,,
\end{equation}
\begin{equation}\label{f6}
f_6=-\frac{2}{G^4}M^{2'}+\frac{2\left(1-M^2\right)\left(F-2\right)}{RG^4}
\,,
\end{equation}
\begin{equation}\label{f7}
f_7=\frac{2}{R^2 G^2} M^{2'}-\frac{\left(1-M^2\right)\left(F-2\right)\left(F+4\right)}{2R^3G^2}-\frac{F}{R}f_4
\,,
\end{equation}
\begin{equation}\label{f8}
f_8=-\frac{F-2}{R} f_5
\,,
\end{equation}
\begin{equation}\label{f9}
f_9=\frac{2}{M^2}\left(Q-\frac{\nu^2}{2R^2}\right)
\,.
\end{equation}

%\newpage
\section[]{physical quantities and differential equations of model}
The MHD integrals have the following forms, 
\begin{equation}\label{psi}
\Psi_{A}=\sqrt{4\pi\rho_{\star}\left(1+\delta \alpha+\mu \delta_{0}
\alpha^{\epsilon}\right)}\,,
\end{equation}
\begin{equation}\label{omega}
\Omega=\frac{V_{\star}}{r_{\star}}\sqrt{\frac{\mu \alpha^{\epsilon-1}+\xi}{
1+\delta \alpha+\mu \delta_{0} \alpha^{\epsilon}}}\,,
\end{equation}
\begin{equation}\label{L}
L=V_{\star} r_{\star} \sqrt{\frac{\mu \alpha^{\epsilon+1}+\xi \alpha^2 }{
1+\delta \alpha+\mu \delta_{0} \alpha^{\epsilon}}}\,.
\end{equation}
The three ordinary differential equations for the functions of $R$ are
\begin{equation}\label{arxdiaf}
\left.
\begin{array}{r}
f_0^{'}-f_6-f_9=0 \\
f_4^{'}-f_7+\xi \left(f_5^{'}-f_8 \right) -\delta f_9 =0 \\
\mu \left(\frac{f_5^{'}}{\epsilon}-f_8-\delta_0 f_9 \right)=0
\end{array}
\right\} 
\end{equation}
or using the definitions of $P_0\,,P_1$ and $F$
\begin{equation}\label{dg2dr}
\begin{array}{l}
\displaystyle{
{d G^2\over dR} = - {{F-2}\over R}G^2}
\,,
\end{array}
\end{equation}

\begin{eqnarray}\label{dfdr}
\begin{array}{l}
\displaystyle{
\frac{dF}{dR}=\frac{F}{1-M^2}\frac{d M^2}{dR}-\frac{F\left(F-2
\right)}{2R}-}
\\ \\
\displaystyle{
\frac{F^{2}-4}{2R \left(1-M^2 \right)}-
\frac{2G^2RP_{1}}{1-M^2 }-}
\\ \\
\displaystyle{
\frac{2\xi R}{M^2 \left(1-M^2 \right)^{3}}\left[\left(2M^2-1\right)
G^4-M^4+2M^2\left(1-G^2\right)\right]}
\end{array}
\end{eqnarray}

\begin{eqnarray}\label{dm2dr}
\begin{array}{l}
\displaystyle{
\frac{d M^2}{dR} = \frac{M^2 \left(1-M^2 \right)}{ \left(2M^2-1
\right) G^4-M^4}\left\{
2 \epsilon \delta_0 G^2 \left(1-M^2 \right) 
\left(Q-\frac{\nu^{2}}{2 R^2}\right)+ \right.}
\\ \\
\displaystyle{
\left.
\frac{F-2}{R}\left[\left(\epsilon+1\right)M^2-\left(\epsilon-1\right)
G^4\right]\right\}}
\end{array}
\end{eqnarray}

\begin{eqnarray}\label{dp1dr}
\begin{array}{l}
\displaystyle{
\frac{dP_{1}}{dR}=- \left[\frac{F^{2}-4}{2R^{2}G^2}+2\xi
\frac{\left(1-G^2\right)^
{2}}{G^2\left(1-M^2 \right)^{3}}\right] \frac{d M^2}{dR}-}
\\ \\
\displaystyle{
\frac{M^2 F}{2R^{2} G^2}\frac{dF}{dR}+
\frac{2\delta}{M^2} \left(Q-\frac{\nu^{2}}{2 R^2}\right)-
\frac{M^2 \left(F^{2}-4\right)\left(F-4\right)}{4R^{3}G^2}+}
\\ \\
\displaystyle{
\xi\frac{\left(F-2\right)\left[\left(2M^2-1\right)
 G^4-M^4\right]}{RG^2M^2\left(1-M^2 \right)^{2}}}
\end{array}
\end{eqnarray}

\begin{eqnarray}\label{dp0dr}
\begin{array}{l}
\displaystyle{
\frac{dP_{0}}{dR}=-\frac{2}{G^4}\frac{d M^2}{dR}+\frac{2}{M^2}
\left(Q-\frac{\nu^{2}}{2 R^2}\right)-
\frac{2M^2\left(F-2\right)}{RG^4}}
\end{array}
\end{eqnarray}
The pressure component $P_2(R)$ is given explicitly in terms of 
the other variables:
\begin{eqnarray}\label{p2}
P_{2}=\frac{\mu}{G^2}\left[\frac{G^4-M^2}{\epsilon M^2\left(1-M^2
\right)}- \left(\frac{1-G^2}{1-M^2}\right)^{2}\right]
\,.
\end{eqnarray}
The functional form of the pressure, Eq. (\ref{pressure}),  
corresponds to the following functional form for the heating function 
\begin{eqnarray}\label{}
\frac{q}{\rho V_r}=\frac{V_{\star}^2}{2 r_{\star}} 
\frac{ {\cal Q}_0 + {\cal Q}_1\alpha + {\cal Q}_2 \alpha^{\epsilon}
}{1+\delta \alpha+\mu \delta_0 \alpha^{\epsilon}}
\end{eqnarray}
with 
${\cal Q}_i=M^{-2\left(\Gamma-1\right)}
\displaystyle{\frac{d}{dR}\left(\frac{M^{2 \Gamma} P_i}{\Gamma-1} \right)}
\,,i=0,1,2$. As discussed in Sec. 2.1, 
one could proceed in the reverse way, i.e., to start with the functional 
form of the heating function and deduce the functional form of  
the pressure Eq. (\ref{pressure}).

\bsp

\label{lastpage}

\end{document}